\documentclass[a4paper,11pt]{article}
\usepackage[utf8]{inputenc}
\usepackage{lmodern}
\usepackage[T1]{fontenc}
\usepackage{amsmath,amsthm,amsfonts,amssymb,bm,mathtools}
\usepackage{dsfont}     % indicator function
\usepackage{enumitem}
\usepackage{graphicx}
\usepackage{bbm}
\usepackage{booktabs}
\usepackage{xcolor}
\usepackage{authblk}
\usepackage[authoryear]{natbib}
\usepackage[right=2.5cm,left=2.5cm,top=2cm,bottom=3cm]{geometry}
\usepackage[font={small,it}]{caption} % the options makes the font italic in floats
\usepackage[colorlinks,linkcolor=blue,citecolor=blue,urlcolor=blue]{hyperref}
\usepackage{subfig}
\usepackage{mathtools}

\usepackage{multirow}
\usepackage{makecell} 
\usepackage{afterpage}

\usepackage{color}
\usepackage[normalem]{ulem}
\definecolor{orange}{RGB}{216,101,0}

\DeclareMathOperator*{\argmax}{arg\,max}

\title{ \textsc{A latent variable model for identifying and characterizing food adulteration} }
\author[1]{Alessandro Casa}
\author[2]{Thomas Brendan Murphy}
\author[2]{Michael Fop}
\affil[1]{Faculty of Economics and Management, Free University of Bozen-Bolzano}
\affil[2]{School of Mathematics and Statistics, University College Dublin }
\date{}                     %% if you don't need date to appear

\begin{document}
\maketitle

\begin{abstract}
Recently, growing consumer awareness of food quality and sustainability has led to a rising demand for effective food authentication methods. Vibrational spectroscopy techniques have emerged as a promising tool for collecting large volumes of data to detect food adulteration. However, spectroscopic data pose significant challenges from a statistical viewpoint, highlighting the need for more sophisticated modeling strategies. To address these challenges, in this work we propose a latent variable model specifically tailored for food adulterant detection, while accommodating the features of spectral data. Our proposal offers greater granularity with respect to existing approaches, since it does not only identify adulterated samples but also estimates the level of adulteration, and detects the spectral regions most affected by the adulterant. Consequently, the methodology offers deeper insights, and could facilitate the development of portable and faster instruments for efficient data collection in food authenticity studies. The method is applied to both synthetic and real honey mid-infrared spectroscopy data, delivering precise estimates of the adulteration level and accurately identifying which portions of the spectra are most impacted by the adulterant.
\end{abstract}

\section{Introduction}\label{sec:introduction}

In the dynamic food production system landscape, we are witnessing a growing awareness of food quality, traceability, and sustainability among consumers, retailers, and food processors. In this context, the issue of food adulteration is attracting significant attention due to its economic implications, association with fraudulent practices, and its potential detrimental effects on public health and the environment  \citep{hassoun2020fraud,dimitrakopoulou2023does}. Food adulteration is defined as the act of intentionally degrading the quality of food by adding or replacing food substances with cheaper alternatives, or by removing some components; however, food authenticity can also refer to the history of production or geographical origin \citep{TEENTEH2014265}. Many different foods are susceptible to adulteration, however the most frequently adulterated foods are typically the more expensive ones, or those whose production levels are heavily influenced by external factors such as weather and harvest conditions.

As a consequence, food authenticity studies, aiming to determine if foodstuffs are what they are claimed to be, are increasingly important. \citet{reid2006recent} provide a review of the analytical strategies that are the most widely adopted in these studies. The authors point out that frequently employed methodologies often require costly and time-consuming laboratory extraction processes for data collection, which sometimes compromise their effectiveness. In contrast, vibrational spectroscopy techniques, such as Fourier transform near-infrared (NIR) and mid-infrared (MIR) spectroscopy, are taking on a more central role \citep{sun2009infrared}. In fact, these techniques are nondisruptive and require little sample preparation. As such, they represent a relatively cheap and rapid alternative to collect large volumes of data on different food samples. From a practical standpoint, when MIR spectroscopy is employed to analyze a material, light is directed through a sample of that material at different wavelengths in the mid-infrared range. The interaction between the light and the sample activates the chemical bonds in the sample, leading to the absorption of light energy. The quantity of absorbed energy, measured at different wavelengths, produces the spectrum of the analyzed sample. Spectra have been shown to be a source of valuable information; consequently, these methods are currently being used to analyze different materials in different fields, which span from medicine to astronomy.

In the context of food authenticity, spectroscopic data are increasingly regarded as a central tool, since each spectrum can be seen as a distinctive fingerprint of the chemical composition of a sample \citep[see, e.g.,][]{downey1996authentication, downey1997near,connolly2006spectroscopic,toher2007comparison, kamal2015analytical}. A key principle in interpreting such fingerprints is the \emph{Beer–Lambert law} \citep{beer1852bestimmung}, which provides the intuition that the intensity of absorption at a given wavelength is directly linked to the concentration of the corresponding chemical component. In the case of adulteration, this means that the observed spectrum lies on a continuum between that of the pure food and that of the pure adulterant, depending on their relative proportions in the combination of spectra.

Building on the intuition that an adulterated sample's spectrum is a mixture of pure and adulterant components, we frame the food authentication problem within the class of \emph{individual-level mixture models} \citep{pritchard2000inference, erosheva2002grade, heller2008statistical, airoldi2014handbook, gruhl2013tale, hou-liu2022}. These models can be regarded as a relaxation of the more traditional \emph{population-level mixture models} \citep[see][for recent reviews]{mcnicholas2016mixture, bouveyron2019model}, widely employed for clustering and classification. In population-level mixtures, each unit is assumed to belong to a single group, with a one-to-one correspondence between groups and mixture components. By contrast, individual-level mixtures allow units to simultaneously belong to multiple components with varying degrees of membership. This feature is particularly well suited to the applied context, given the potentially large variability in adulteration levels across food samples. While population-level mixture-based classification and clustering methods have already been explored in this setting \citep[see, e.g.,][]{dean2006using, murphy2010variable,Fop2022}, they provide only a coarse discrimination between adulterated and non-adulterated samples. As a consequence, such models tend to identify only heavily adulterated samples, while grouping the remaining ones into a single class. On the other hand, individual-level mixtures enable the development of more refined statistical tools, capable of offering higher resolution, and directly estimating the degree of adulteration in each sample. Notably, these models have found applications well beyond food authentication, ranging from document analysis \citep{blei2003latent} to the ranking of political candidates \citep{gormleyMurphy} and the modeling of survey responses \citep{erosheva2007describing}.

While the individual-level mixture framework is conceptually well-suited for the food authentication task, its practical implementation must account for the challenging nature of spectroscopic data. In fact, from a statistical viewpoint, these data pose several challenges that must be addressed in the modeling steps. First, spectral datasets are typically high-dimensional, as absorbance is measured over hundreds or thousands of wavelengths, each representing an observed variable. Moreover, chemical processes often induce strong and complex correlations among wavelengths. The information in spectral data is known to be spread across different spectral regions and structured in a compounded way, often depending on the analyzed biological material \citep{casa2022parsimonious}.

To address these challenges in the context of food authentication, we propose a sparse individual-level mixture model tailored to high-dimensional spectral data. The introduced model not only enables the identification of adulterated food samples, but also gives richer insights by detecting the percentage of the adulterant in the sample. Moreover, a convenient parsimonious representation of the relationships between wavelengths is provided. Lastly, the estimation procedure identifies which spectral regions are impacted by the adulterant; identifying these regions could lead to the development of cheaper and faster portable instruments to collect data for food authenticity studies.

The paper is structured as follows. In Section \ref{sec:model_def}, we introduce our proposal, outlining various modeling approaches designed to address the peculiar characteristics of the considered application. In Section \ref{sec:modelEst}, we outline the model estimation strategy, with a focus on model selection steps and initialization procedures. An exploration of the performances on synthetic data is reported in Section \ref{sec:Simulations}, while in Section \ref{sec:realDataAnalysis} we report the results obtained on honey mid-infrared spectral data. Section \ref{sec:Conclusion} ends the paper with some final remarks and discussing interesting avenues for future research. 

\section{Sparse individual-level mixture model}\label{sec:model_def}

As introduced in Section~\ref{sec:introduction}, the Beer-Lambert law implies that each observed spectrum is an additive mixture of the spectra of its constituents, namely the pure/non-adulterated food and the pure adulterant. Intuitively, if a sample contains a fraction $g$ of adulterant and $(1-g)$ of pure food, its absorbance at each wavelength can be described as a linear combination of the two corresponding spectra. In this sense, the observed spectrum represents a continuum between the two extremes: the pure food when $g = 0$, and the pure adulterant when $g = 1$.

Building on this intuition, we adopt a probabilistic formulation for the observed spectra. Let $\mathbf{Y} = \{\mathbf{y}_1, \dots, \mathbf{y}_n \}$ denote the collection of spectra, with $\mathbf{y}_i \in \mathbb{R}^p$, and let $g_i^{\texttt{A}} \in [0,1]$ represent the level of adulteration for the $i$-th sample, $i = 1, \dots, n$. Moreover, we denote with $\bm\delta = (\delta_1, \dots, \delta_p)$ the \emph{mean-shifts} vector, which quantify the difference in absorption between the pure food and the adulterant, whose mean spectra are $\bm\mu^{\texttt{P}} = (\mu_1, \dots, \mu_p)$ and $\bm\mu^{\texttt{A}} =(\mu_1 + \delta_1, \dots, \mu_p + \delta_p)$, respectively. The expected value of $\mathbf{y}_i$ can be expressed as a convex combination of its constituents
\begin{align}
\mathbb{E}(\mathbf{y}_i \vert g_i^\texttt{A}) & = (1 - g_i^\texttt{A})\bm\mu^{\texttt{P}} + g_i^\texttt{A}\bm\mu^{\texttt{A}} \nonumber \\
& = \bm\mu^{\texttt{P}} + g_i^\texttt{A}(\bm\mu^{\texttt{A}} - \bm\mu^{\texttt{P}}) = \bm\mu^{\texttt{P}} + g_i^\texttt{A}\bm\delta . \label{eq:MeanBeerLambert}
\end{align}
To capture variability in the observed measurements, we assume that $\mathbf{y}_i$ follows a multivariate Gaussian distribution with mean \eqref{eq:MeanBeerLambert}, such as
\begin{eqnarray}\label{eq:firstMod}
	(\mathbf{y}_i | \mathbf{g}^{\texttt{A}}, \bm\Theta) \sim \mathcal{N}_p(\bm\mu^{\texttt{P}} + \bm\delta g_{i}^{\texttt{A}}, \bm\Sigma) ,
\end{eqnarray}
where $\bm\Theta = \{ \bm\mu^{\texttt{P}}, \bm\delta, \bm\Sigma \}$ and $\mathbf{g}^{\texttt{A}} = (g_1^{\texttt{A}}, \dots, g_n^{\texttt{A}})$, while $\bm\Sigma$ accounts for measurement noise, natural variability, and dependence between the spectral regions. 

Interestingly, model \eqref{eq:firstMod} can be regarded as a special case of a partial membership model \citep[PMM,][]{heller2008statistical}, which in turn can be viewed as a specific formulation of the broader class of individual-level mixture models. In particular, it corresponds to a PMM with two components, the pure food and the pure adulterant, where the adulteration level $g_i^\texttt{A}$ plays the role of the partial membership weight. The key distinction from a general PMM lies in the additive mean structure in \eqref{eq:MeanBeerLambert}, which is chemically motivated by the Beer–Lambert law. This connection provides a rigorous statistical framing of our proposal within the class of individual-level mixtures. As a result, the model inherits the advantages of individual-level over population-level mixtures, enabling a finer characterization of adulteration, explicitly quantifying the degree of membership to the adulterant component.

Practically, to fit the model \eqref{eq:firstMod} and determine the regions of the spectrum relevant to distinguish between pure and adulterated samples, we develop a sparsity-inducing procedure aimed at maximizing the following penalized log-likelihood
\begin{eqnarray}\label{eq:penLik}
	\ell_p(\mathbf{g}^{\texttt{A}}, \bm\Theta; \mathbf{Y}) = \sum_{i = 1}^n \log \phi(\mathbf{y}_i; \bm\mu^{\texttt{P}} + \bm\delta g_{i}^{\texttt{A}}, \bm\Omega) - p_{\bm\lambda}(\mathbf{g}^{\texttt{A}}, \bm\delta, \bm\Omega) ,
\end{eqnarray}
where $\phi(\cdot; \bm\mu^{\texttt{P}} + \bm\delta g_{i}^{\texttt{A}}, \bm\Omega)$ denotes the density of a multivariate Gaussian distribution with mean vector $\bm\mu^{\texttt{P}} + \bm\delta g_{i}^{\texttt{A}}$ and precision matrix $\bm\Omega = \bm\Sigma^{-1}$. 
In addition, the second term in \eqref{eq:penLik} represents a generic penalty function, governed by hyperparameter vector $\bm\lambda$, which determines its magnitude. The choice of this penalty is guided by the specific modeling objectives. In this work, we consider
\begin{eqnarray}\label{eq:penalty}
	p_{\bm\lambda}(\mathbf{g}^{\texttt{A}}, \bm\delta, \bm\Omega) = \lambda_{\text{g}} \Vert \mathbf{g}^{\texttt{A}} \Vert_1 + \lambda_\delta \Vert \mathbf{D}\bm\delta \Vert_1 + \lambda_\Omega \Vert \bm\Omega \Vert_1 ,
\end{eqnarray}
where $\bm\lambda = (\lambda_g, \lambda_\delta, \lambda_\Omega)$ is the vector of the hyperparameters controlling the strength of the penalization on $\mathbf{g}^{\texttt{A}}, \bm\delta$ and $\bm\Omega$ respectively, and $\Vert \cdot \Vert_1$ represents the $L_1$-norm, with $\Vert \mathbf{A} \Vert_1 = \sum_{jh} \lvert A_{jh}\rvert$. 

The specific choice of the penalty is motivated by the application considered in this work. In fact, the third term in \eqref{eq:penalty} corresponds to a graphical lasso penalty on the precision matrix $\bm\Omega$ \citep{friedman:etal:2008}. By shrinking some elements of $\bm\Omega$ exactly to zero, this approach helps mitigate the difficulties that arise when estimating full covariance or precision matrices in high-dimensional scenarios; since the number of parameters grows quadratically with $p$, direct estimation can  be infeasible or produce unreliable results when $p$ is large. On the other hand, sparse precision matrices, thanks to the connection with Gaussian graphical models \citep[GGM,][]{whittaker:1990}, lend themselves to a convenient visual interpretation in terms of conditional independence. Zero entries in $\bm\Omega$ denote missing edges in the corresponding graph, thus implying that two variables are independent conditioning on all the others features. This approach appears particularly suitable, as data explorations and insights from the literature suggest that conditional dependencies among wavelengths are typically sparse.

The first term in \eqref{eq:penalty} adds a $L_1$-penalty on the levels of adulteration $g_i^{\texttt{A}}$, for $i = 1, \dots, n$. By setting exactly to zero some of the $g_i^{\texttt{A}}$, this penalty allows to obtain more parsimonious solutions and to build a first coarse classification procedure. In fact, $g_i^{\texttt{A}} = 0$ implies that the $i$-th sample has not been adulterated, thus discriminating between pure food and adulterated samples. Considering an $L_1$-penalty, the proposal automatically implements a thresholding procedure, with the threshold being a function of the hyperparameter $\lambda_g$, which needs to be tuned (see Section \ref{sec:modelSelection} for details about the tuning). Nonetheless, if user-defined thresholds are more appropriate for a specific application, for example based on domain-specific or applied considerations, one can simply switch off the automatic shrinkage by setting $\lambda_g = 0$ and manually threshold the resulting non-sparse estimates of the adulterant levels.

The second term in \eqref{eq:penalty} refers to the generalized lasso class of problems \citep{gen_lasso}. Generalized lasso encompasses all those situations where the $L_1$-penalty is considered to impose structural constraints, which could include, but are not limited to, sparsity on some coefficients. A careful specification of the matrix $\mathbf{D}$ could lead to different well-known penalized schemes such as lasso \citep{tibshirani1996regression}, fused lasso \citep{tibshirani2005sparsity} and trend filtering \citep{kim_TF,trend_filtering}. In this work, $\mathbf{D}$ is selected in order to impose a sparse fused lasso penalty on $\bm\delta$. Therefore $\mathbf{D}^\top = [ \tilde{\mathbf{D}}^\top \, \vert \, \mathbb{I}_p] \in \mathbb{R}^{ p \times (2p-1)}$ with 
\begin{eqnarray*}
\tilde{\mathbf{D}} = 
\begin{bmatrix}
-1 & 1 & 0 & \dots & 0 & 0\\
0 & -1 & 1 & \dots & 0 & 0 \\
\vdots & \vdots & \vdots & \ddots & \vdots & \vdots\\
0 & 0 & 0 & \dots & -1 & 1 \\
\end{bmatrix} .
\end{eqnarray*}
This amounts to penalize the discrete first derivative, since the \emph{j}-th row of $\tilde{\mathbf{D}}$ enforces a penalization on the absolute difference among adjacent coefficients $\lvert \delta_{j+1} - \delta_{j} \rvert$. Also, by stacking the identity matrix $\mathbb{I}_p$ after $\tilde{\mathbf{D}}$, a standard lasso penalty is imposed element-wise on the mean-shifts $\delta_j$, for $j = 1,\dots, p$. This specific choice of $\mathbf{D}$ is driven by the need to construct a parsimonious modeling framework capable of handling large dimensional scenarios, as the ones arising in spectroscopy. From a practical standpoint, the sparse fused lasso penalty automatically takes into account that adjacent wavelengths tend to behave similarly, and that some spectral regions are likely not to be influenced by the presence of the adulterant. By shrinking to zero pairwise differences $\lvert \delta_{j + 1} - \delta_{j} \rvert$, it is possible to automatically detect portions of the spectrum whose behavior is influenced by the same underlying chemical processes. Moreover, inducing element-wise sparsity, it allows to perform variable selection which could lead to richer insights and to a more comprehensive interpretation of the adulteration effects. 

This specification of $\mathbf{D}$ is not binding, as other choices are possible. If no information about local behavior is available, simple lasso penalty on $\bm\delta$ may be used. Otherwise, linear and polynomial trend filtering can be considered by specifying $\mathbf{D}$ such that second or higher-order derivatives are penalized (see \citet{gen_lasso} or \citet{ferraccioli2023adaptive} for an application to functional linear regression). 

\section{Model estimation}\label{sec:modelEst}

We outline the procedure to estimate the parameters in $\bm\Theta$ and the adulterant membership vector $\mathbf{g}^{\texttt{A}}$, treated as an unknown vector of parameters. In realistic food authentication applications, the food being analyzed is usually known; therefore, $\bm\mu^{\texttt{P}}$ is either known in advance or estimated arbitrarily well by obtaining spectra from a large number of pure food samples. Consequently, hereafter we consider the centered data $\mathbf{y}^{c}_i = \mathbf{y}_i - \bm\mu^{\texttt{P}}$, for $i = 1,\dots, n$. Estimates for the remaining parameters $\bm\delta, \bm\Omega$ and $\mathbf{g}^{\texttt{A}}$ are then obtained by maximizing the penalized log-likelihood \eqref{eq:penLik} for centered data. Maximization iterates over three partial optimization steps, each targeting one of the aforementioned parameters, as described below.

At the $(t+1)$-th iteration, given the current estimates $\hat{\bm\Omega}^{(t)}$ and $\hat{\mathbf{g}}^{\texttt{A}, (t)}$, the estimation of $\bm\delta$ is carried out relying on the alternating direction method of multipliers \citep[ADMM,][]{boydADMM}, already been used in similar frameworks \citep[see, e.g.,][]{ramdasADMM}. ADMM algorithm effectively solves constrained problems of the form 
\begin{eqnarray}\label{eq:admm_general}
\text{minimize}&& \hspace{0.5cm} f(\bm\alpha) + g(\bm\beta) \\
\text{subject to}&& \hspace{0.45cm} \nonumber \mathbf{A}\bm\alpha + \mathbf{B}\bm\beta = \mathbf{c}.
\end{eqnarray}
This strategy is particularly useful in penalized estimation settings, where faster and easier optimization is often possible by resorting to a variable splitting scheme. We recast the maximization of \eqref{eq:penLik} with respect to $\bm\delta$ as follows 
\begin{eqnarray}\label{eq:admm_our}
\text{minimize}&& \hspace{0.5cm} - \sum_{i = 1}^n \log \phi(\mathbf{y}^c_i; \bm\delta g_{i}^{\texttt{A}}, \bm\Omega) + \lambda_\delta \Vert \bm\gamma \Vert_1  \\
\text{subject to}&& \hspace{0.45cm} \nonumber \mathbf{D}\bm\delta - \bm\gamma = 0 .
\end{eqnarray}
ADMM obtains the estimates by minimizing the augmented Lagrangian of \eqref{eq:admm_our} iteratively, given the separability of the objective function. The closed-form updates, whose derivation is reported in Appendices~\ref{appendixA1}, \ref{appendixA2}, and \ref{appendixA3}, are as follows:
\begin{eqnarray}
	\hat{\bm\delta}^{(s +1)} & = & \left[\hat{\bm\Omega}^{(t)} \sum_{i = 1}^n \left(\hat{g}^{\texttt{A}, (t)}_i\right)^2 + \rho \mathbf{D}^\top \mathbf{D}\right]^{-1} \left[\left(\sum_{i = 1}^n \hat{g}^{\texttt{A}, (t)}_i \hat{\bm\Omega}^{(t)} \mathbf{y}_i^c \right) + \rho \mathbf{D}^\top \hat{\bm\gamma}^{(s)} - \mathbf{D}^\top \hat{\mathbf{u}}^{(s)} \right] \label{eq:deltaADMM}, \\ [8pt]
	\hat{\bm\gamma}^{(s+1)}_j & = & \frac{1}{\rho} \mathcal{S}\left(\hat u_j^{(s)} + \rho \left[\mathbf{D}\hat{\bm\delta}^{(s+1)}\right]_j; \, \lambda_\delta\right) \hspace{0.6cm} \text{for} \,\, j = 1, \dots, 2p-1\label{eq:gammaADMM},\\ [8pt]
	\hat{\mathbf{u}}^{(s+1)} & = & \hat{\mathbf{u}}^{(s)} + \rho\left(\mathbf{D}\hat{\bm\delta}^{(s+1)} - \hat{\bm\gamma}^{(s+1)}\right), \label{eq:dualvariableUpdate}
\end{eqnarray}
where \eqref{eq:dualvariableUpdate} is the dual variable update, $\rho > 0$ is the augmented Lagrangian parameter, and $[\mathbf{D}\hat{\bm\delta}^{(s+1)}]_j$ represents the generic $j$-th element of the vector $\mathbf{D}\hat{\bm\delta}^{(s+1)}$. Moreover $\mathcal{S}(x;\lambda) = \text{sign}(x)(\lvert x \rvert - \lambda)_+$, with $x_+ = \text{max}\{0, x\}$, is the soft-thresholding operator. The choice of $\rho$ could impact the convergence speed \citep[see][for an in-depth discussion]{boydADMM,ramdasADMM}; in this work, we follow the suggestions in \citet[Section 3.4.1,][]{boydADMM} where a dynamic update scheme is proposed to improve convergence in practice and reduce sensitivity to the initial choice of $\rho$.  

In the ADMM framework, different convergence results have been studied \citep[see,][for a comprehensive examination]{boydADMM}. In this work, the stopping criterion is given by conditions on the \emph{dual residuals} $d$, and the \emph{primal residuals} $r$ which, at the $(s+1)$-th iteration, are defined as 
\begin{eqnarray*}
d^{(s+1)} &=& \rho\mathbf{D}^\top\mathbb{I}^{-}(\hat{\bm\gamma}^{(s+1)} - \hat{\bm\gamma}^{(s)}), \\
r^{(s+1)} &=& \mathbf{D}\hat{\bm\delta}^{(s+1)} - \hat{\bm\gamma}^{(s+1)},
\end{eqnarray*}
with $\mathbb{I}^{-}$ being a matrix with elements on the diagonal equal to $-1$ and 0 elsewhere. Specifically, the algorithm is stopped at iteration $s^*$ if $\Vert d^{(s^*)} \Vert_2< \epsilon_d $ and $\Vert r^{(s^*)} \Vert_2< \epsilon_r$, with $\epsilon_d$ and $\epsilon_r$ defined as in \citet[Section 3.3.1,][]{boydADMM}. Lastly, if the stopping criterion is met at a given iteration $s^*$, we denote $\hat{\bm\delta}^{(t+1)} = \hat{\bm\delta}^{s^*}$ the estimate of the mean-shifts vector at the $(t+1)$-th iteration of the global algorithm. Note that, when the global algorithm has reached convergence, the final estimate of the mean-shift for the $j$-th variable is given by $\hat{\delta}_j\mathbbm{1}(\hat\gamma_{(j)} \neq 0)$, where $\mathbbm{1}(\hat\gamma_{(j)} \neq 0) = 1$ if $\hat\gamma_{(j)} \neq 0$ and $0$ otherwise, where $\hat\gamma_{(j)}$ denotes the subset of entries of $\bm\gamma$ corresponding to the rows of $\mathbf{D}$ that impose the element-wise $L_1$-penalty on $\delta_j$. This additional adjustment is applied to ensure exact sparsity on $\bm{\hat \delta}$, as the updates in \eqref{eq:deltaADMM} do not directly enforce it. 

The estimate of the precision matrix is computed by maximizing \eqref{eq:penLik} with respect to $\bm\Omega$. Then, $\hat{\bm\Omega}^{(t+1)}$ is obtained by solving the following optimization problem 
\begin{eqnarray}\label{eq:graphLasso}
	\argmax_{\bm\Omega} \;\;\; \log\text{det}(\bm\Omega) - \text{tr}(\mathbf{S}\bm\Omega) - \lambda_\Omega \Vert \bm\Omega \Vert_1 ,
\end{eqnarray}
with $\bm\Omega$ constrained to be positive definite and $\mathbf{S}$ the sample covariance matrix computed as 
$$
	\mathbf{S} = \frac{1}{n}\sum_{i = 1}^n (\mathbf{y}_i^c - \hat{\bm\delta}^{(t+1)}\hat{g}_i^{\texttt{A}, (t)})(\mathbf{y}_i^c - \hat{\bm\delta}^{(t+1)}\hat{g}_i^{\texttt{A}, (t)})^\top .
$$
To solve the problem in \eqref{eq:graphLasso}, we resort to the coordinate descent graphical lasso procedure as proposed in \citet{friedman:etal:2008}. 

The estimate for the adulterant membership vector $\hat{\mathbf{g}}^{\texttt{A}, (t+1)}$ is obtained by taking the derivative of \eqref{eq:penLik} with respect to $g_i^{\texttt{A}}$, for $i = 1, \dots, n$, and equating it to zero. The following solution is then obtained
\begin{eqnarray*}
\hat g_i^{\texttt{A}, (t+1)} = \frac{1}{(\hat{\bm\delta}^{(t+1)})^\top \hat{\bm\Omega}^{(t+1)}\hat{\bm\delta}^{(t+1)}}\left[ (\hat{\bm\delta}^{(t+1)})^\top\hat{\bm\Omega}^{(t+1)}\mathbf{y}_i^c - \lambda_g \tau_i\right] ,
\end{eqnarray*}
with $\tau_i$ being the sub-gradient of $\lvert g_i^{\texttt{A}} \rvert$, defined as
\begin{eqnarray*}
\tau_i = \begin{cases}
\hspace*{0.28cm} 1 \hspace*{1cm} \text{if} \; g_i^{\texttt{A}} > 0 \\
-1 \hspace*{1cm} \text{if} \; g_i^{\texttt{A}} < 0 \\
\hspace*{0.3cm} 0 \hspace*{1cm} \text{if} \; g_i^{\texttt{A}} = 0  ,
\end{cases}
\end{eqnarray*}
for $i = 1, \dots, n$. The estimate for $g_i^{\texttt{A}, (t+1)}$ can be expressed more concisely in terms of the soft-thresholding operator as follows
\begin{eqnarray}
\label{eq:solutionG}
\hat g_i^{\texttt{A}, (t+1)} = \frac{1}{(\hat{\bm\delta}^{(t+1)})^\top \hat{\bm\Omega}^{(t+1)}\hat{\bm\delta}^{(t+1)}} \mathcal{S}\left( (\hat{\bm\delta}^{(t+1)})^\top\hat{\bm\Omega}^{(t+1)}\mathbf{y}_i^c; \lambda_g\right) ,
\end{eqnarray}
and readers can refer to the Appendix \ref{appendixA4} for the analytical derivation. The estimates are constrained to lie between 0 and 0.5, implicitly assuming that the maximum level of adulteration is equal to 0.5. Taking steps from the implementation in \texttt{glmnet} package \citep{glmnetPack}, this is obtained by considering an additional thresholding step where 
\begin{eqnarray}
\label{eq:thresholdG}
\hat g_i^{\texttt{A}, (t+1)} = \begin{cases}
0 \hspace*{1.97cm} \text{if} \; \hat g_i^{\texttt{A}, (t+1)} \le 0 \\
\hat g_i^{\texttt{A}, (t+1)} \hspace*{1cm} \text{if} \; 0< \hat g_i^{\texttt{A}, (t+1)} < 0.5 \\
0.5 \hspace*{1.72cm} \text{if} \;\hat g_i^{\texttt{A}, (t+1)} \ge 0.5 .
\end{cases}
\end{eqnarray}
Lastly, the procedure easily allows to consider some elements of $\hat{\mathbf{g}}^{\texttt{A}}$ as known, thus incorporating some degree of supervision in the modeling framework (see Section \ref{sec:supervision} for more details). 

The three partial optimization steps described above are repeated until a convergence criterion is met. In this work, convergence is achieved when the relative improvement in the objective function \eqref{eq:penLik} between subsequent iterations is smaller than a predefined threshold.

\subsection{Model selection}\label{sec:modelSelection}
The estimation procedure in Section \ref{sec:modelEst} considers $\bm\lambda = (\lambda_g, \lambda_\delta, \lambda_\Omega)$ as fixed. However, in practical applications $\bm\lambda$ is not known in advance, and it has to be selected resorting to model selection tools. Here, we select $\lambda_g, \lambda_\delta$ and $\lambda_\Omega$ by means of the Bayesian information criterion \citep[BIC,][]{schwarz1978estimating}, which is widely used in a broad range of applications where standard finite mixture models are employed \citep[see e.g.,][and references therein]{fraley2002model}. Coherently with the penalized model-based clustering literature \citep[see][]{pan2007penalized,casa2022group}, and relying on previously obtained results \citep{Zou2007a,lian2011shrinkage}, we employ a modified version of the BIC which reads as follows
\begin{eqnarray}\label{eq:BIC}
\text{BIC} = 2\log L(\hat{\mathbf{g}}^{\texttt{A}}, \hat{\bm\Theta}) - \nu_0\log(n) ,
\end{eqnarray} 
where $\log L(\hat{\mathbf{g}}^{\texttt{A}}, \hat{\bm\Theta})$ is the log-likelihood evaluated at $(\hat{\mathbf{g}}^{\texttt{A}}, \hat{\bm\Theta})$, while $\nu_0$ is the number of parameters not shrunk to 0 in the estimation phase. Within the sparse fused lasso framework, the number of free parameters in the vector $\hat{\bm\delta}$ is counted following the suggestions in \citet{tibshirani2005sparsity}.

Operationally, a sequential conditional hyperparameter optimization scheme can be adopted, where the elements of $\bm\lambda$ are optimized one at a time, while keeping the other elements fixed to their values from the previous iteration. Specifically, at the $v$-th iteration, given $\bm\lambda^{(v)} = (\lambda_g^{(v)}, \lambda_\delta^{(v)},\lambda_\Omega^{(v)})$, the value of $\lambda_\delta^{(v + 1)}$ is selected by choosing from a grid of values, according to the corresponding BIC values computed as in \eqref{eq:BIC}, and with $\lambda_\delta^{(v)},\lambda_\Omega^{(v)}$ fixed. The same procedure is applied for the other hyperparameters, conditioning on the most recently updated values. The search scheme ends when the hyperparameters stop updating. Although this conditional search may provide suboptimal results, it avoids the computational burden of exhaustive grid searches. The required number of operations is reduced from $\mathcal{O}(n_{\text{gr}}^3)$, where $n_{\text{gr}}$ is the length of the grid for each hyperparameter, to $\mathcal{O}(3n_{\text{gr}}n_{\text{it}})$, where $n_{\text{it}}$ is the number of iterations until the search converges. This results in fewer operations, provided that $n_{\text{it}} < n_{\text{gr}}^2/3$, a condition typically satisfied in practical experiments. 

\subsection{Initialization procedure}\label{sec:initializationProc}
Initialization strategies might play a crucial role in accelerating the convergence of the algorithm introduced in Section \ref{sec:modelEst} and in improving its performance. To find reasonable initial values $\hat{\bm\delta}^{(0)}$ and $\hat{\mathbf{g}}^{\texttt{A}, (0)}$, we propose an ad-hoc heuristic initialization strategy. 

The procedure builds on the following modification of the model in \eqref{eq:firstMod}
\begin{eqnarray*}
\mathbf{y}_i = \bm\mu^{\texttt{P}} + g_{i}^{\texttt{A}}\bm\delta + \bm\varepsilon_i, \;\;\;\;\; i = 1, \dots, n ,
\end{eqnarray*}
where $\bm\varepsilon_i \in \mathbb{R}^p$ is a residual term with $\mathbb{E}(\bm\varepsilon_i) = \mathbf{0}$ and $\mathbb{E}(\bm\varepsilon_i^2) = \sigma^2\mathbb{I}_p$, while all the other quantities are defined as before. This amounts to considering a linear regression model, relying solely on conditions on the first two moments of $\bm\varepsilon_i$, thus avoiding distributional assumptions.
Consequently, the initial values for the parameters are obtained by minimizing 
\begin{eqnarray}\label{eq:optimiz}
\text{RSS} = \sum_{i =1}^n \sum_{j = 1}^p (y_{ij} - \mu^{\texttt{P}}_j - g_{i}^{\texttt{A}}\delta_j)^2 ,
\end{eqnarray}
with respect to  $\bm\delta$ and $g_{i}^{\texttt{A}}$, with $\bm\mu^{\texttt{P}}$ again considered as known. Parameters are then estimated by means of an iterative procedure, which, at the $(t+1)$-th iteration,  alternates the following two steps
\begin{enumerate}
	\item Given $\hat{g}_i^{\texttt{A},(t)}$, obtain  
	$$
	\hat{\delta}_{j}^{(t+1)} = \frac{\sum_{i = 1}^n \left( y_{ij} - \mu^{\texttt{P}}_j\right)\hat{g}_i^{\texttt{A},(t)}}{\sum_{i = 1}^n \left(\hat{g}_i^{\texttt{A},(t)}\right)^2}, \;\;\;\;\; j = 1, \dots, p
	$$ 
	\item Given $\hat{\bm\delta}^{(t+1)}$, obtain $\hat{g}_i^{\texttt{A},(t+1)}$, for $i = 1,\dots, n$, by optimizing \eqref{eq:optimiz}. 
\end{enumerate}
This heuristic procedure disregards the correlation structure among the variables. As such, it does not provide a starting value for the precision matrix $\bm\Omega$, which can initially be set to be a diagonal matrix. This initialization strategy has been shown to provide promising initial values in practical applications, both for a non-sparse $\bm\delta$ and for the vector  $\mathbf{g}^{\texttt{A}}$. Lastly, in those cases where the mean vector for the pure food needs to be estimated, the procedure can be easily extended to find reasonable initial values for $\bm\mu^{\texttt{P}}$. 

\subsection{Different degrees of supervision}\label{sec:supervision}
In the considered application, it seems reasonable to recast the estimation procedure in terms of a semi-supervised framework. In fact, food authenticity studies are often conducted in controlled experimental environments, with knowledge about the analyzed food and the adulterant to detect; from a statistical viewpoint, this amounts to assuming the availability of some labeled spectra in the observed sample. 

Let $n^{\texttt{P}}$ and $n^{\texttt{A}}$ denote, respectively, the number of labeled spectra from pure food and adulterated samples for which $g_i^{\texttt{A}}$ is known. Depending on the experimental conditions, various configurations for $(n^{\texttt{P}}, n^{\texttt{A}})$ are possible. The number of known adulterated and pure samples can vary: both could be greater than zero, or alternatively, only $n^{\texttt{P}} \neq 0$ or $n^{\texttt{A}} \neq 0$. These different combinations introduce different types of semi-supervision and amounts of information, which is then exploited in the estimation step. 

In the next sections, different situations are explored by means of true and synthetic data. More specifically, in Section \ref{sec:Simulations} $n^{\texttt{P}}$ and $n^{\texttt{A}}$ are fixed to predetermined percentages of the sample size $n$, while in Section \ref{sec:realDataAnalysis} distinct scenarios are investigated, to showcase the impact of different degrees of supervision on the quality of the estimates. These explorations may therefore provide initial guidance on how to set up the experiments in order to ease authentication tasks in a food adulteration framework.

\section{Simulated data experiments}\label{sec:Simulations}
In this section, we evaluate the performance of the method on synthetic data generated from model \eqref{eq:firstMod}, considering various configurations of $n$, $p$,  membership vectors $\mathbf{g}^{\texttt{A}}$ and mean-shifts vectors $\bm\delta$. Specifically, two shapes for $\bm\delta$ are considered, in the following referred to as $\bm\delta^{(1)}$ and $\bm\delta^{(2)}$. The first one is a variation of the Mexican hat function, generally defined as
$$ f(x) = 4(1 - x^2)\text{exp}\{-x^2/2\} \quad \text{for } x \in [-5, 5], $$
and here adjusted through the following truncation process
$$ 
\bm\delta^{(1)}(x) = 
\begin{cases}
    2.4  &\;\; \text{ if } f(x) > 2.4 \, ,\\
    -1.6 &\;\; \text{ if } f(x) < -1.6 \, , \\
    0 \, &\;\; \text{ if } f(x) \in [-0.4, 0] \, , \\
    f(x) &\;\; \text{ otherwise}, 
\end{cases} 
$$
which modifies the function so that it exhibits a locally flat signal while being zeroed out in other parts of the domain. 
In this scenario, $\bm\mu^{\texttt{P}}$ has been set equal to 0. 
In the second setting $\bm\delta^{(2)}$ is generated as a perturbation of $\bm\mu^{\texttt{P}}$, here defined as a piecewise cubic function in $[0, 1]$ constructed from a B-spline basis with $15$ equispaced internal knots. The perturbation is obtained by modifying a subset of spline coefficients generating the curve.

Precision matrix $\bm\Omega = \text{diag}(\omega_1, \dots, \omega_p)$ is considered as fixed for all the scenarios presented in the following subsections, since its recovery does not represent the main aim of the study. Throughout the simulations, the percentage of adulterated observations, for which $g_i^{\texttt{A}} \neq 0$, is fixed and equal to $30$\% of the total sample size $n$. Moreover, consistent with the considerations in Section \ref{sec:supervision}, we consider $20\%$ of the observations as labeled, proportionally split between adulterated and pure.

The simulation study has multiple goals. Firstly, we want to test if the proposal is able to identify whether the observations have been adulterated or not. This amounts to a classification task, assessing the ability to correctly estimate $\hat g_i^{\texttt{A}} = 0$ when $g_i^{\texttt{A}} = 0$. As assessment criteria, we resort to classification-oriented measures such as accuracy, sensitivity and specificity. In addition, we monitor the quality of the estimation of the membership vector by looking at the \emph{mean absolute error} (MAE) defined as 
\begin{eqnarray}\label{eq:mae}
    \text{MAE} = \frac{1}{n}\sum_{i = 1}^n \vert \hat{g}_i^{\texttt{A}} - g_i^{\texttt{A}} \vert \, .
\end{eqnarray}
Moreover, to evaluate the quality of the estimates of the mean-shifts vector $\bm\delta$, we consider the \emph{mean squared error} (MSE) 
\begin{eqnarray}\label{eq:mse}
     \text{MSE} = \frac{1}{p}\sum_{j = 1}^p (\hat\delta_j - \delta_j)^2 \, .
\end{eqnarray}
Our last objective is to evaluate the performance of variable selection, monitoring whether the method can accurately identify which variables are affected by the adulterant. This amounts to checking which $\hat\delta_j$ have been estimated as zero and comparing them with the actual $\delta_j = 0$, for $j = 1, \dots, p$. As for the membership vector, we rely on accuracy, sensitivity and specificity. In both these cases, the \emph{true positives} are considered to be the elements correctly shrunk to zero, the \emph{true negatives} the ones correctly estimated as different from zero, and the \emph{false positives} and \emph{false negatives} as the ones wrongly shrunk to zero or wrongly estimated as different from zero, respectively.

The following analyses are implemented within the \texttt{R} environment \citep{R}. Parameters are initialized according to the procedure outlined in Section \ref{sec:initializationProc}, while the hyperparameter vector $\bm\lambda$ is tuned according to the sequential strategy outlined in Section \ref{sec:modelSelection}. For all the different simulated scenarios $B = 100$ Monte Carlo samples have been generated; results are reported in the next subsections.  

\subsection{Scenario 1}\label{sec:simScenario1}
In this section, samples have been generated for distinct combinations of $n = \{ 100, 250, 500 \}$ and $p = \{ 50, 100, 150\}$. Mean-shifts vector has been set to $\bm\delta^{(1)}$ and the levels of adulteration $g_i^{\texttt{A}}$ randomly sampled from $\{0.1, 0.2, 0.3 \}$ for $i \in \mathcal{I}$, with $\mathcal{I} = \{i : g_i^{\texttt{A}} \neq 0 \}$. The aim is to investigate the impact of $n$ and $p$ on the measures introduced in the previous section; results are reported in Table \ref{tab:resScen1}. 

\begin{table}[t]
\caption{On the left: mean absolute error (MAE, multiplied by $10^2$), accuracy (ac$_g$), sensitivity (sn$_g$) and specificity (sp$_g$), for the estimation of $\mathbf{g}^{\texttt{A}}$. On the right: mean squared error (MSE, multiplied by $10^2$), accuracy (ac$_\delta$), sensitivity (sn$_\delta$) and specificity (sp$_\delta$), for the estimation of $\bm\delta$. Results are reported for different combinations of $n$ and $p$.}
\label{tab:resScen1}
\centering
\addtolength{\tabcolsep}{-1pt}
\resizebox{\textwidth}{!}{%
\begin{tabular}{lccc||ccc}
\multicolumn{1}{c |}{} & \multicolumn{3}{c||}{\textbf{Adulteration recovery}} & \multicolumn{3}{c}{\textbf{Signal recovery}} \\
\hline
\hline
 \multicolumn{1}{c|}{} & $n = 100$ & $n = 250$ & $n = 500$ & $n = 100$ & $n = 250$ & $n = 500$\\ \hline
 \multicolumn{1}{l|}{\multirow{3}{*}{$p = 50$}} & MAE = 0.537  & MAE = 0.460 & MAE = 0.515 & MSE = 1.151 & MSE = 0.431 & MSE = 0.105 \\
\multicolumn{1}{l|}{} & ac$_g$ = 0.909 & ac$_g$ = 0.909 & ac$_g$ = 0.914  & ac$_\delta$ = 0.915  & ac$_\delta$ = 0.936 & ac$_\delta$ = 0.931\\
\multicolumn{1}{l|}{} & sn$_g$ = 0.866  & sn$_g$ = 0.870 & sn$_g$ = 0.879 & sn$_\delta$ = 1.000  & sn$_\delta$ = 1.000 & sn$_\delta$ = 1.000 \\
\multicolumn{1}{l|}{} & sp$_g$ = 1.000 & sp$_g$ = 1.000 & sp$_g$ = 1.000 & sp$_\delta$ = 0.879 & sp$_\delta$ = 0.908 & sp$_\delta$ = 0.899 \\
\hline
\multicolumn{1}{c|}{\multirow{3}{*}{$p = 100$}} & MAE = 0.395  & MAE = 0.292 & MAE = 0.300 & MSE = 1.204   & MSE = 0.443 & MSE = 0.122  \\
\multicolumn{1}{l|}{} & ac$_g$ = 0.920  & ac$_g$ = 0.926  & ac$_g$ = 0.876 & ac$_\delta$ = 0.927   & ac$_\delta$ = 0.968 & ac$_\delta$ = 0.939 \\
\multicolumn{1}{l|}{} & sn$_g$ = 0.883   & sn$_g$ = 0.894 & sn$_g$ = 0.825 & sn$_\delta$ = 0.996  & sn$_\delta$ = 0.999  & sn$_\delta$ =  1.000 \\
\multicolumn{1}{l|}{} & sp$_g$ = 1.000  & sp$_g$ = 1.000 & sp$_g$ = 1.000  & sp$_\delta$ = 0.895 & sp$_\delta$ =  0.949 & sp$_\delta$ = 0.906 \\
\hline
\multicolumn{1}{c|}{\multirow{3}{*}{$p = 150$}} & MAE = 0.389 & MAE = 0.246 & MAE = 0.250 & MSE = 1.175 & MSE = 0.471  & MSE = 0.171 \\
\multicolumn{1}{l|}{} & ac$_g$ = 0.918& ac$_g$ = 0.932 & ac$_g$ = 0.922 & ac$_\delta$ = 0.930  & ac$_\delta$ = 0.976  & ac$_\delta$ = 0.967 \\
\multicolumn{1}{l|}{} & sn$_g$ = 0.880 & sn$_g$ = 0.904 & sn$_g$ = 0.890  & sn$_\delta$ = 0.991 & sn$_\delta$ = 0.997 & sn$_\delta$ = 0.999 \\
\multicolumn{1}{l|}{} & sp$_g$ = 1.000 & sp$_g$ = 1.000 & sp$_g$ = 1.000 & sp$_\delta$ = 0.902 & sp$_\delta$ = 0.964 & sp$_\delta$ = 0.949 \\
 \hline
\end{tabular}
}
\end{table}

Across all configurations, the percentage of the adulterant is always well recovered, with average absolute errors between 0.002 and 0.005. MAE values seem not to be dependent on the different sample sizes, while more pronounced fluctuations occur when changing $p$, with better results when the number of variables is larger. Changes in $n$ and $p$ seem not to have strong impact on the performance of the classification metrics, which attain satisfactory values regardless of the sample size and the number of variables. A thorough analysis of the values of the sensitivity and the specificity suggests that the method does not produce false positives, with a slightly larger inclination to obtain false negatives. This means that there are no adulterated samples wrongly identified as pure, while some of the pure ones are not correctly estimated as such. In food authenticity, where the adulterant might be dangerous for human health, this over-conservative behavior is preferable, as it would enable subsequent testing to detect potential adulterants in cases where false negatives occur.

In terms of signal recovery, the quality of the estimates of $\bm\delta$ seems to be sensitive to changes in the sample size, while different values of $p$ do not strongly affect the results. In particular, increasing $n$ leads to improvements in the MSE values, highlighting that higher informative contents imply more precise estimation of $\bm\delta$. On the other hand, classification metrics are less impacted by changes in $n$ and $p$, as they attain extremely good values for all the configurations. This seems to suggest that the accurate identification of the regions of the spectrum affected by the adulterant is a simpler task than precisely estimating the impact of the adulterant itself. Differently from the adulteration recovery, here the methodology produces more false positives and almost no false negatives; this serves as an indication that the procedure tends to over-shrink some of the $\delta_j$'s. Note that potential over-shrinking also likely affects the overall quality of the estimate $\hat{\bm\delta}$ by inducing a source of bias for the $\delta_j \neq 0$.  

\subsection{Scenario 2}\label{sec:simScenario2}
We generate samples for both $\bm \delta^{(1)}$ and $\bm \delta^{(2)}$, with $g_i^{\texttt{A}} = \{0.1, 0.2, 0.3 \}$ for $i \in \mathcal{I}$, $n = 250$ and $p = 100$. In particular, for both $\bm \delta^{(1)}$ and $\bm \delta^{(2)}$, we explore the \emph{weak signal} and \emph{strong signal} setting. When considering $\bm \delta^{(1)}$ the signal is weakened or strengthened by considering as mean-shifts vectors $\bm \delta^{(1)}/2$ or $1.5 \bm \delta^{(1)}$ respectively. On the other hand, for $\bm \delta^{(2)}$ the intensity of the signal is tuned by changing the amount of spline coefficients modified when generating the curve, as described in Section \ref{sec:Simulations}; this allows to obtain a different number of $\delta^{(2)}_j \neq 0$, for $j = 1, \dots, p$. Specifically, for the \emph{weak signal} case on average the 25.4\% of the variables are impacted by the adulterant, while for the \emph{strong signal} case the number grows to 35.4\%. These scenarios allow to explore situations where the concept of strength of the adulterant changes; in fact, for $\bm \delta^{(1)}$ the strength is depending on inflation or deflation of the signal, while for $\bm \delta^{(2)}$ it depends on the size of the portion of the domain impacted by the adulterant. The aim is to evaluate how different types and strengths of adulterant effects influence the ability to detect adulteration and accurately recover the signal. Results are displayed in Table \ref{tab:resScen2}.

\begin{table}[t]
\caption{Cf. Table \ref{tab:resScen1}. Results refer to Scenario 2. Different intensities of the signal (strong and weak) are considered for both $\bm \delta^{(1)}$ and $\bm \delta^{(2)}$. Values of MAE and MSE are multiplied by $10^2$.}
\label{tab:resScen2}
\centering
\addtolength{\tabcolsep}{-1pt}
\begin{tabular}{lcc||cc}
\multicolumn{1}{c |}{} & \multicolumn{2}{c||}{\textbf{Adulteration recovery}} & \multicolumn{2}{c}{\textbf{Signal recovery}} \\
\hline
\hline
 \multicolumn{1}{c|}{} & Strong Signal & Weak Signal & Strong Signal & Weak Signal \\ \hline
 \multicolumn{1}{l|}{\multirow{3}{*}{$\bm \delta^{(1)}$}} & MAE =  0.196  & MAE = 0.638  & MSE = 0.486  & MSE = 0.456  \\
\multicolumn{1}{l|}{} & ac$_g$ = 0.928  & ac$_g$ = 0.918 & ac$_\delta$ = 0.961   & ac$_\delta$ = 0.967 \\
\multicolumn{1}{l|}{} & sn$_g$ = 0.897 & sn$_g$ = 0.883 & sn$_\delta$ = 1.000  & sn$_\delta$ = 0.988 \\
\multicolumn{1}{l|}{} & sp$_g$ = 1.000 & sp$_g$ = 1.000 & sp$_\delta$ = 0.949 & sp$_\delta$ = 0.955 \\
\hline
\multicolumn{1}{c|}{\multirow{3}{*}{$\bm \delta^{(2)}$}} & MAE = 0.259  & MAE = 0.569 & MSE = 0.254 & MSE = 0.168 \\
\multicolumn{1}{l|}{} & ac$_g$ =  0.914 & ac$_g$ = 0.922  & ac$_\delta$ = 0.942  & ac$_\delta$ =  0.941\\
\multicolumn{1}{l|}{} & sn$_g$ = 0.876  & sn$_g$ = 0.888 & sn$_\delta$ = 0.936 & sn$_\delta$ =  0.940 \\
\multicolumn{1}{l|}{} & sp$_g$ = 1.000 & sp$_g$ = 0.999  & sp$_\delta$ = 0.951 & sp$_\delta$ =  0.952\\
\hline
\end{tabular}
\end{table}

As expected, it stands out how the precise estimation of the percentage of the adulterant is strongly influenced by the intensity of the signal for both $\bm \delta^{(1)}$ and $\bm \delta^{(2)}$; in fact, in both cases the MAE values improves significantly in the strong signal scenarios. However, even considering weak signals, these values range from 0.0019 to 0.0063 highlighting an overall satisfactory recovery of the adulteration. This is confirmed also by the classification metrics, which show that the proposal is capable of discriminating between the pure and the adulterated samples. Some of the conclusion drawn in Section \ref{sec:simScenario1} are still valid here: in fact, specificity values are still nearly always equal to 1, while most of the errors are false negatives. Note that, the results in terms of classification are much less influenced by the strength of the signal. 

The intensity of the signal seems to have a less clear impact in terms of signal recovery. In fact, while the MSE values seem satisfactory for all the scenarios, when considering $\bm \delta^{(2)}$ the mean-shifts vector is estimated better when considering weaker signal; this could be linked to the way the signal has been weakened, with fewer elements of $\delta_j^{(2)} \neq 0$. In this setting if the elements different from zero are correctly identified, as it appears to be from the classification metrics, the MSE would automatically tend to decrease. Generally speaking, classification metrics attain good results and provide indications different from the ones previously obtained. In fact, the over-shrinkage witnessed in Section \ref{sec:simScenario1} is not evident anymore, with more heterogeneous types of error, especially for $\bm \delta^{(2)}$. However, results still seem insensitive to changes in the strength of the signal, once again demonstrating the adequacy of the proposal for variable selection.

\subsection{Scenario 3}\label{sec:simScenario3}
We fix mean-shifts vector to $\bm \delta^{(1)}$, considering both the weak and the strong signal case, as described in Section \ref{sec:simScenario2}. The sample size is set equal to $n = 250$ and the number of variables $p = 100$. On the other hand, we consider two different settings for the membership vector:
\begin{itemize}
    \item \emph{Medium adulteration}, where $g_i^{\texttt{A}} = 0.2$, for $i \in \mathcal{I}$;
    \item \emph{Weak adulteration}, where $g_i^{\texttt{A}} = 0.05$, for $i \in \mathcal{I}$. 
\end{itemize}
The objective is to evaluate how different levels of adulteration impact the results, and how they interact with the strength of the adulterant. Results are reported in Table \ref{tab:resScen3}.

\begin{table}[t]
\caption{Cf. Table \ref{tab:resScen1}. Results refer to Scenario 3. Different combinations of strength of the adulterant effects (strong and weak) and levels of adulteration (weak and medium) are considered. Values of MAE and MSE are multiplied by $10^2$.}
\label{tab:resScen3}
\centering
\addtolength{\tabcolsep}{-1pt}
\begin{tabular}{lcc||cc}
\multicolumn{1}{c |}{} & \multicolumn{2}{c||}{\textbf{Adulteration recovery}} & \multicolumn{2}{c}{\textbf{Signal recovery}} \\
\hline
\hline
 \multicolumn{1}{c|}{} & Weak Ad & Medium Ad & Weak Ad & Medium Ad \\ \hline
 \multicolumn{1}{l|}{\multirow{3}{*}{Weak signal}} & MAE = 1.477  & MAE = 0.621 & MSE = 24.149 & MSE = 0.430 \\
\multicolumn{1}{l|}{} & ac$_g$ = 0.712 & ac$_g$ = 0.918 & ac$_\delta$ = 0.720  & ac$_\delta$ = 0.965  \\
\multicolumn{1}{l|}{} & sn$_g$ = 0.997  & sn$_g$ = 0.881 & sn$_\delta$ = 0.627  & sn$_\delta$ = 0.988  \\
\multicolumn{1}{l|}{} & sp$_g$ = 0.073 & sp$_g$ = 1.000 & sp$_\delta$ = 0.997 & sp$_\delta$ = 0.951  \\
\hline
\multicolumn{1}{c|}{\multirow{3}{*}{Strong signal}} & MAE = 0.274  & MAE = 0.182  & MSE = 19.920 & MSE = 0.508 \\
\multicolumn{1}{l|}{} & ac$_g$ =  0.927 & ac$_g$ = 0.925 & ac$_\delta$ =  0.977 & ac$_\delta$ = 0.969 \\
\multicolumn{1}{l|}{} & sn$_g$ = 0.895 & sn$_g$ = 0.892  & sn$_\delta$ = 0.977  & sn$_\delta$ = 1.000  \\
\multicolumn{1}{l|}{} & sp$_g$ = 1.000  & sp$_g$ = 1.000  & sp$_\delta$ = 0.978 & sp$_\delta$ = 0.950 \\
\hline
\end{tabular}
\end{table}

A first look shows clearly the interplay between the intensity of the signal and the percentage of adulteration. In the setting where both the adulteration and the signal are weak, the method fails to adequately estimate the parameters $\mathbf{g}^{\texttt{A}}$ and $\bm\delta$; in fact, the errors in terms of MAE and MSE are substantially higher with respect to all the previously explored settings. Moreover, classification performances essentially show how the proposal fails to discriminate between pure and adulterated samples, and between variables impacted or not by the adulterant. A closer inspection suggests that the penalty tends to set to zero most of the $g_i^{\texttt{A}}$, not identifying the small amounts of adulterant in the samples. In terms of signal recovery, the method seems not to be able to shrink to zero the $\delta_j$'s not influenced by the adulterant. The performance rapidly improves as soon as either strong signal or medium adulteration are considered. In these situations, our proposal tends to behave again as discussed in the previous sections. Here, the mean absolute error when estimating the percentage of the adulterant ranges from 0.0018 to 0.0062 and the classification metrics, both for adulteration and signal recovery, shows adequate performances. This confirms that, as soon as the effect of the adulterant is strong enough, the procedure is capable of identifying adulterated samples and to perform variable selection, even in cases with small percentages of adulteration. Lastly, further investigations may be needed to understand why, with strong signal and weak adulteration, the method fails to adequately estimate the $\bm\delta$, as shown by the high value of the MSE. A closer look to the result seems to suggest that, even if the $\delta_j = 0$ are correctly identified, the signal is strongly over-shrunk for those $\delta_j \neq 0$.

\section{Application to honey MIR spectroscopy data}\label{sec:realDataAnalysis}
\subsection{Data description}
\begin{figure}[t]
    \centering
    \includegraphics[scale = 0.6]{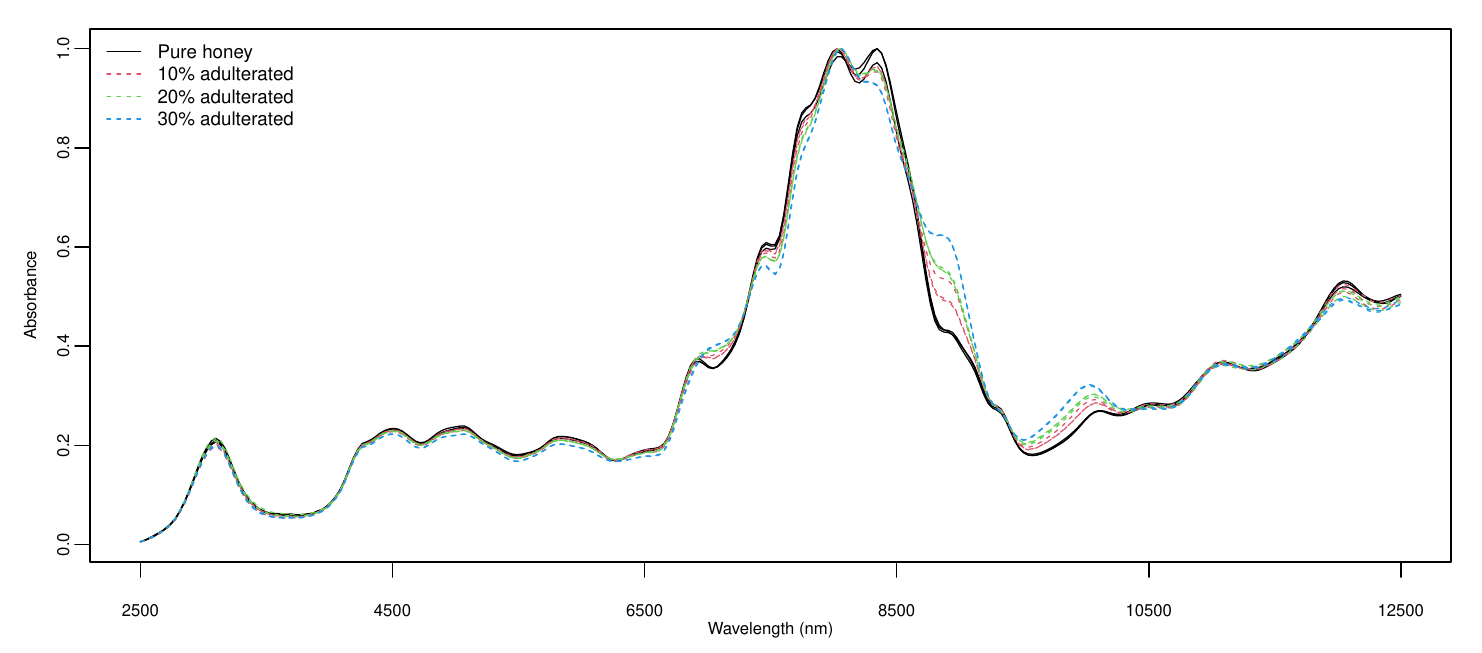}
    \caption{Spectra recorded for a subset of honey samples with different adulteration levels. A random selection of curves is shown for each adulteration group.}
    \label{fig:exampleSpectra}
\end{figure}

In this section, the developed food authentication methodology is applied to honey MIR spectral data. The dataset includes MIR spectra of Irish artisanal honey, encompassing both authentic samples and those adulterated with five types of sugar syrups; detailed information regarding the data collection process can be found in \citet{kelly2006application}. In this work, we focus on a single adulterant, namely {\em beet sucrose solution}. Honey samples have been collected directly from beekeepers in Ireland at various time points between 2000 and 2003. Pure honey samples have been subsequently adulterated in batches of $40$, each with varying levels of beet sucrose adulteration. The resulting dataset comprises a total of $n = 410$ MIR spectra, including $n_H = 290$ from pure honey and $n_B = 120$ spectra from honey adulterated with beet sucrose at three different levels ($10\%, 20\%, 30\%$). The availability of the information on adulteration levels is essential for evaluating the ability of the proposed method to accurately estimate the percentage of adulterant in each sample. The spectra comprise absorbance values measured at $285$  wavelengths in the mid-infrared region, spanning $2500$ to $12500$~nm. Figure \ref{fig:exampleSpectra} provides a graphical representation of a subset of the data. A visual inspection of the spectra behavior reveals that beet sucrose affects only specific portions of the spectral domain. This observation underscores the importance of analytical tools capable of identifying the most relevant wavelengths in the context of food authentication.  

To reflect typical experimental settings (see Section~\ref{sec:supervision}), we assume partial supervision and treat 10\% of the samples as labeled, proportionally split between pure and adulterated. Moreover, we consider the centered data $\mathbf{y}^{c}_i$, for $i = 1, \dots, n$, with $\bm\mu^{\texttt{P}}$ considered as known. In line with recommendations from the literature \citep[see, e.g.,][]{murphy2010variable}, a variable aggregation step was carried out prior to conducting the analysis to reduce the computational burden. In fact, when analyzing spectral data, the strong correlations between wavelengths make it possible to consider slightly lower resolutions while preserving the majority of the informative content. Consequently, we aggregate pairs of adjacent wavelengths, thus considering $p = 143$ features. Finally, when applying the estimation procedure described in Section \ref{sec:modelEst}, the hyperparameter vector $\bm\lambda = (\lambda_g, \lambda_\delta, \lambda_\Omega)$ is tuned over a predefined grid of values. The optimal $\bm\lambda$ is selected using the sequential optimization scheme outlined in Section \ref{sec:modelSelection}. 

\begin{figure}[t]
    \centering
    \includegraphics[scale = 0.8]{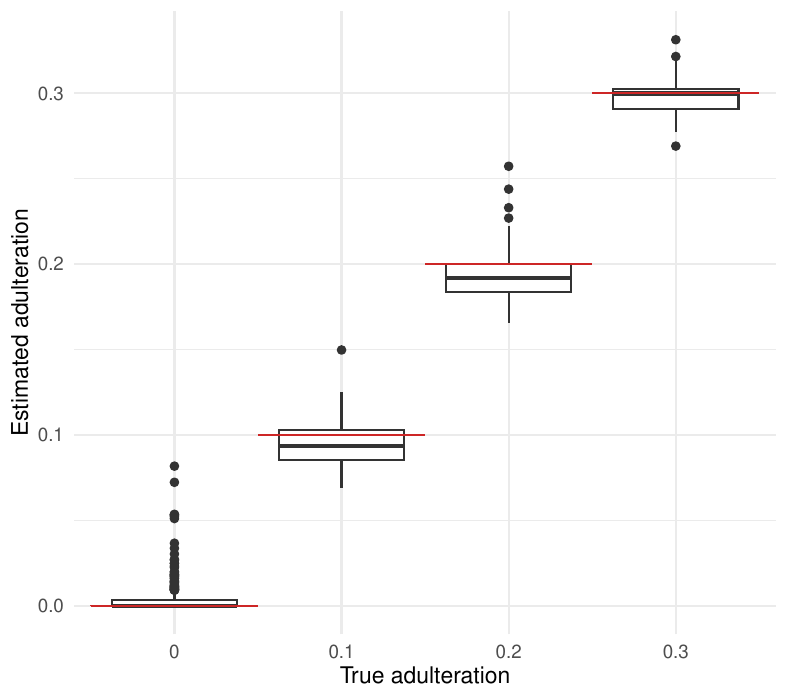}
    \caption{Boxplots of the estimated adulteration levels $\hat{g}_i^{\texttt{A}}$ grouped by the true adulteration level $g_i$, shown on the x-axis and highlighted with the red horizontal segments.}
    \label{fig:boxplotContam}
\end{figure}

\subsection{Adulteration level estimation}
Figure \ref{fig:boxplotContam} compares $\hat{g}_i^{\texttt{A}}$, the estimated adulteration levels, and $g_i^{\texttt{A}}$, the true adulteration levels, for $i = 1,\dots,n$. The proposed method demonstrates excellent performance in recovering beet-sucrose contamination, with a slight underestimation of $g_i^{\texttt{A}}$ for adulterated samples. This behavior might be attributed to the shrinkage effect of the $L_1$-penalty imposed on $g_i$ in \eqref{eq:penalty}, which in turn effectively drives some of the $\hat{g}_i^{\texttt{A}}$ values to exactly zero for pure samples. In fact, the $62.45\%$ of the pure honey samples are correctly identified as non-adulterated, while the remaining pure samples anyway exhibit $\hat{g}_i^{\texttt{A}}$ values lower than those of the adulterated samples. More specifically, in terms of classification metrics we obtain $\text{ac}_g = 0.734$, $\text{sn}_g = 0.624$ and $\text{sp}_g = 1$. Lastly, we obtain $\text{MAE} = 0.007$, further underscoring the effectiveness of our approach, not only in determining whether a sample has been adulterated but also in accurately estimating the specific level of adulteration, with an average estimation error of approximately $0.7\%$.  These insights are crucial, as they enable a reliable assessment of adulteration, with significant implications on production systems, food safety, and potentially public health.

\subsection{Signal recovery and wavelength selection}
\begin{figure}[t]
    \centering
    \includegraphics[scale = 0.6]{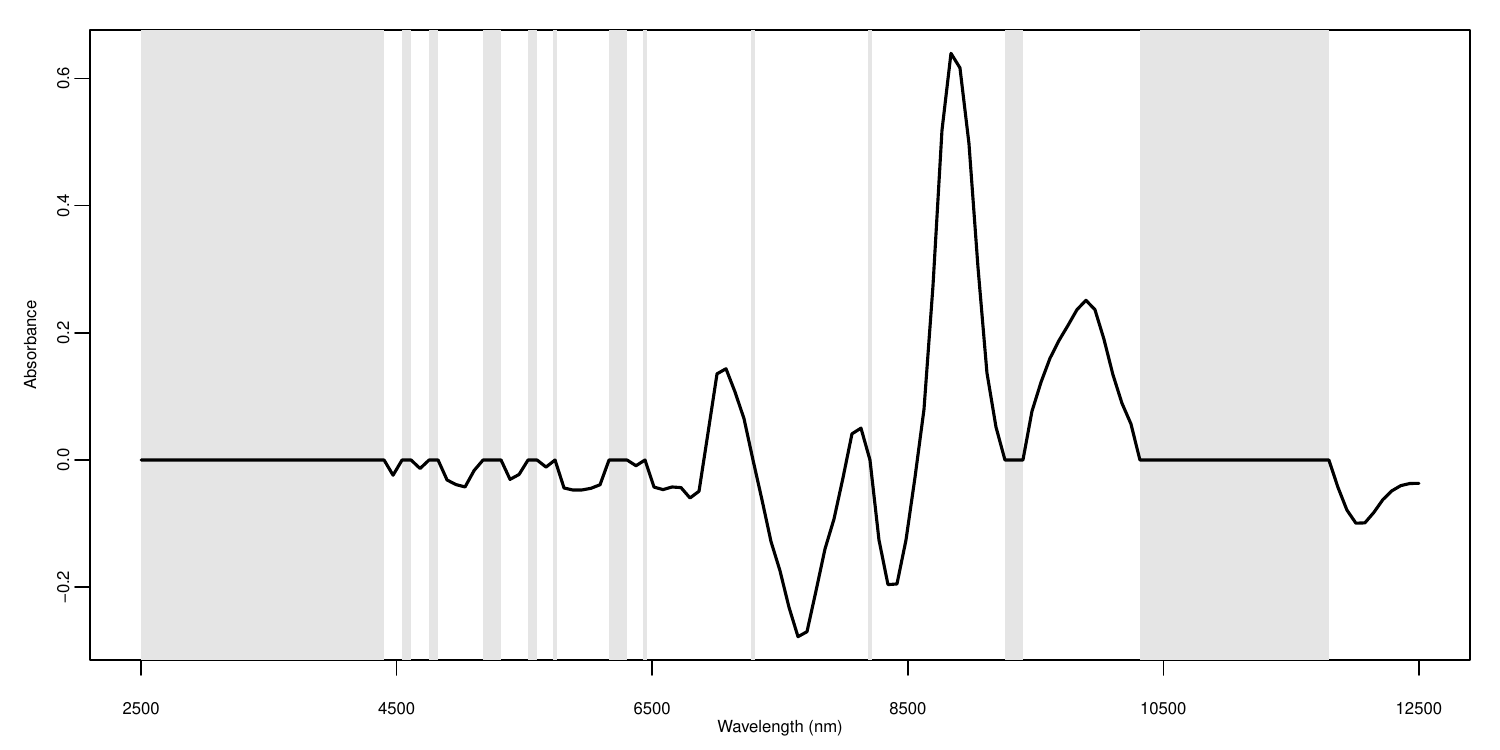}
    \caption{Estimated mean-shifts vector $\hat{\bm\delta}$ (black line) across the wavelength range. The grey-shaded areas denote wavelengths where $\hat\delta_j$ is zero.}
    \label{fig:estMeanShift}
\end{figure}

Figure \ref{fig:estMeanShift} graphically presents the variable selection results, showing the estimated mean-shifts vector $\hat{\bm{\delta}}$ with shaded regions indicating wavelengths where $\hat{\delta}_j = 0$. The model clearly suggests that several spectral regions have not been affected by the adulteration process, which aligns with preliminary observations available from the visual inspection of the raw spectra in Figure \ref{fig:exampleSpectra}.
Chemical interpretation provides further validation. Previous studies have indicated that the spectral region most influenced by the sugar composition in honey lies between $6800$ and $11500$~nm \citep[see, e.g.,][]{hineno1977infrared,kelly2006potential}. Building on this, \citet{kelly2006application} evaluated models for detecting food adulteration by focusing their analyses on this specific wavelength range. These considerations align with the estimate shown in Figure \ref{fig:estMeanShift}, where the higher absorption mean-shifts are located within the aforementioned region. A key advantage of the developed method lies in its ability to identify this region in a data-driven manner, without excluding other spectral regions prior to the analysis. Further insights can be drawn by comparing our results with previous findings in literature. In fact, it has been pointed out \citep{hineno1977infrared} that bands between $6800$ and $8700$~nm are usually connected with bending modes of C-C-H, C-O-H and O-C-H groups. Moreover, the peaks visible in Figure \ref{fig:exampleSpectra}, between $8700$ and $11000$~nm are linked with C-H and C-C stretching modes, with the second one potentially due to O-H vibrations \citep{vasko1972infrared}. While these attributes are characteristics of honey samples, \citet{kelly2006application} note that differences between pure and adulterated honey are likely to manifest around these regions; this seems to be supported by the results in Figure \ref{fig:estMeanShift}.
These preliminary observations shed light on the impact of the adulteration process and provide chemical confirmation of the soundness of the proposed method. Further in-depth explorations and closer collaboration with experts in the fields can enrich the analyses, and lead to a deeper understanding of the food adulteration process as a whole, and ultimately contribute to the development of portable spectrometers, focusing only on specific spectral regions. This would make food authentication procedures more cost-effective and accessible to a wider audience, with beneficial impacts from both economic and health perspectives.

\subsection{Influence of supervision level}
To conclude, we briefly explore the impact of different degrees of supervision on the model estimates, as discussed in Section \ref{sec:supervision}. Specifically, we consider three distinct configurations for $(n^{\texttt{P}}, n^{\texttt{A}})$: (1) $n^{\texttt{P}} \neq 0$ and $n^{\texttt{A}} = 0$, with $n^{\texttt{P}}$ set to $10\%$ of the pure spectra; (2) we explore the opposite situation where $n^{\texttt{P}} = 0$ and $n^{\texttt{A}}$ is set to $10\%$ of the total number of adulterated samples; (3) both $n^{\texttt{P}}$ and $n^{\texttt{A}}$ are non-zero but smaller than in the previous analyses, as they are set to the $5\%$ of the pure and adulterated spectra respectively. In all three scenarios, the hyperparameter vector $\bm\lambda$ remains equal to the same as the one tuned for the previous analyzes. This approach is adopted to maintain feasible computational times, implicitly assuming that changes in $(n^{\texttt{P}}, n^{\texttt{A}})$ should not significantly impact the data structure or the optimal value for $\bm\lambda$. 

When only information on pure samples is included in the model ($n^{\texttt{P}} \neq 0, n^{\texttt{A}} = 0$) the performance in recovering $g_i^\texttt{A}$ deteriorates, with an average error of $6.3\%$ ($\text{MAE} = 0.063$). A closer inspection reveals that, in terms of classification of the samples, the method still yields reasonable results, achieving $\text{ac}_g = 0.727$, $\text{sn}_g = 0.601$ and $\text{sp}_g = 1$, with $60.15\%$ of pure samples correctly identified as such. These findings suggest that accurately estimating the adulteration level is a more challenging task with respect to the coarser discrimination between pure and adulterated samples. The more pronounced degradation of the MAE value can be attributed to the model's tendency to estimate $\hat{g}_i^{\texttt{A}} = 0.5$ for the highly adulterated samples, effectively estimating them at the maximum possible adulteration level allowed by the thresholding mechanism in \eqref{eq:thresholdG}. Therefore, when the model lacks prior information on adulterated samples, it identifies them as pure adulterant, thereby failing to accurately determine their actual level of adulteration. 

On the other hand, when $n^{\texttt{P}} = 0$ and $n^{\texttt{A}} \neq 0$, with only information about adulterated samples exploited in the model, results are more similar to the ones obtained when both $n^{\texttt{P}}$ and $n^{\texttt{A}}$ are different from zero. Here, $\text{MAE} = 0.007$ indicating an equally satisfactory retrieval of the adulteration level. In terms of classification, we obtain $\text{ac}_g = 0.761$, $\text{sn}_g = 0.672$ and $\text{sp}_g = 1$, with with $67.24\%$ of pure samples correctly identified. These results represents a slight improvement in classification performance compared to the initial scenario considered in this section. This suggests that the model is better able to distinguish between non-adulterated and adulterated spectra when provided only with information on the latter ones. 

Lastly, when both $n^{\texttt{P}}$ and $n^{\texttt{A}}$ are non-zero, but the level of supervision is reduced to the $5\%$ of the samples, the results show $\text{MAE} = 0.012$, $\text{ac}_g = 0.689$, $\text{sn}_g = 0.561$ and $\text{sp}_g = 1$. This reduction in supervision leads to a decline in the quality of the results, suggesting, as expected, that higher levels of supervision have a positive impact on the accuracy of the model estimates. 

Note that the three differently supervised scenarios have been compared in terms of classification and estimation of the adulteration levels, as no ground truth is available for the mean-shifts $\bm\delta$. However, an examination of $\hat{\bm \delta}$ under the different settings reveals that the estimates are not substantially affected by varying levels of supervision. The only scenario where $\hat{\bm \delta}$ is notably impacted is when $n^{\texttt{P}} \neq 0$ and $n^{\texttt{A}} = 0$; in this case, the overestimation of the adulteration levels, and the interplay between $g_i^{\texttt{A}}$ and $\bm\delta$ causes a visible underestimation of the latter. 

These comparisons, while limited to the analyzed dataset, provide valuable guidance in designing future studies on food authentication.  By highlighting the importance of balanced supervision and its impact on both classification and estimation accuracy, this study underscores the critical role of leveraging information from both pure and adulterated samples, with particular emphasis on the importance of having some degree of prior information on the latter ones. These findings can inform the development of targeted sampling strategies and analytical methodologies, contributing to enhanced detection of food adulteration.

\section{Discussion and conclusion}\label{sec:Conclusion}
In this paper, we present a sparse modification of individual-level mixture models tailored to address the distinct characteristics of spectroscopic data, in the context of food authenticity studies. This is accomplished through a penalized estimation strategy that employs suitably chosen penalties. This approach eases the extraction of parsimonious summaries from complex data structures, often being more interpretable and informative. Notably, our method represents a significant improvement over traditional data-driven techniques for food authentication, which typically rely on simplistic binary classifications of food samples as either adulterated or non-adulterated. In contrast, our approach takes a significant step forward by not only classifying food samples but also enabling precise estimation of the exact percentage of adulteration; this is crucial in situations where accurate assessments of the adulterant is critical, especially in those cases where the adulteration may present serious health risks. Furthermore, the penalized estimation introduces a wavelengths selection mechanism, having two main objectives. First, it accounts for the fact that adjacent wavelengths tend to behave similar, thereby automatically capturing one of the most important behavior in spectroscopy data. Second, it identifies which wavelengths are more affected by adulteration. When combined with subject-matter knowledge, this identification could unveil previously unknown mechanisms and interactions between the pure food and the adulterant from a chemical standpoint. Additionally, it could pave the way for the development of data collection instruments that focus on some specific portions of the spectrum, enabling to gather less noisy data in a faster and more cost-effective manner. 

The proposed method has been successfully tested both on simulated and real honey MIR spectral data. However, several avenues for future research remain open. First, spectroscopic data inherently possess a functional nature. While functional data analysis methods have been somewhat overlooked in this framework \citep{saeys2008potential}, recent studies have demonstrated their promising potential \citep[see, e.g.,][]{codazzi2022gaussian,ferraccioli2023adaptive}. As such, it may be worthwile to consider the functional nature of the data when exploiting partial membership models for food authentication. Interestingly, recent efforts in this direction have been made by \citet{marco2024functional}. Building on this work could be a promising starting point, with modifications to better address the characteristics of spectroscopy data which require special attention also when adopting functional approaches, as shown by \citet{ferraccioli2023adaptive}.

In this work, we have adopted a frequentist approach based on the maximization of a penalized likelihood. However, previous explorations \citep[see][]{casa2023partial} have demonstrated the potential benefits of employing a fully Bayesian framework, and this possibility is still currently under exploration. In this setting, a careful prior elicitation would allow for the incorporation of available knowledge about the data characteristics, while still enabling sparsification of $\bm\delta$. One key advantage of this approach would be the ability to naturally quantify estimation uncertainty. This would provide a more informative understanding through the construction of credible intervals, particularly for the mean-shifts. 

Lastly, the paper has focused on the case where adulteration arises from a single adulterant. Although this is a reasonable assumption in many applied contexts, it may be restrictive in some specific scenarios where little prior knowledge is available. Importantly, the Beer–Lambert law naturally accommodates extensions to multiple adulterants, as the observed spectrum would correspond to a convex combination of several pure spectra. Generalizing the proposed model in this direction could substantially broaden its applicability.

\appendix
\section{Analytical derivations for parameter updates}\label{appendix}
The ADMM algorithm solves the constrained problem in \eqref{eq:admm_our} by iteratively minimizing the corresponding augmented Lagrangian. Following the general formulation in \eqref{eq:admm_general}, we set $\alpha = \delta$ and $\beta = \gamma$. The objective function is separable with 
\begin{eqnarray}
f(\bm\delta) & = &  - \sum_{i = 1}^n \log \phi(\mathbf{y}_i^c; \bm\delta g_i^\texttt{A}, \bm\Omega) \propto \frac{1}{2}\sum_{i = 1}^n (\mathbf{y}_i^c - \bm\delta g_i^\texttt{A})^\top \bm\Omega (\mathbf{y}_i^c - \bm\delta g_i^\texttt{A})\label{eq:f} \\    
g(\bm\gamma) & = &  \lambda_\delta \Vert \bm\gamma \Vert_1 .\label{eq:g}
\end{eqnarray}
The augmented Lagrangian for \eqref{eq:admm_our} is
\begin{eqnarray}\label{eq:augmLagrangian}
    L_\rho(\bm\delta, \bm\gamma, \mathbf{u}) = f(\bm\delta) + g(\bm\gamma) + \mathbf{u}^\top(\mathbf{D}\bm\delta - \bm\gamma) + \frac{\rho}{2} \Vert \mathbf{D}\bm\delta - \bm\gamma \Vert_2^2
\end{eqnarray}
where $\mathbf{u}$ is the vector of dual variables and $\rho$ the augmented Lagrangian parameter as defined in Section \ref{sec:modelEst}. The ADMM updates for $\bm\delta$, $\bm\gamma$ and $\mathbf{u}$ are then obtained by sequentially minimizing \eqref{eq:augmLagrangian} with respect to each variable, as shown in the next subsections. Note that, in the following, we will omit the superscripts to ease readability.

\subsection{Derivation for the $\hat{\bm\delta}$ update} \label{appendixA1}
The update for $\hat{\bm\delta}$ in \eqref{eq:deltaADMM} is obtained by minimizing \eqref{eq:augmLagrangian} with respect to $\bm\delta$. As shown, for example, in \citet{zhu2017augmented}, this is equivalent to solve the following minimization problem 
$$
\hat{\bm\delta} = \arg\min_{\bm\delta} f^*(\bm\delta, \bm\gamma,\mathbf{u}) =  \arg\min_{\bm\delta}  \left( f(\bm\delta) + \frac{\rho}{2} \Vert \mathbf{D}\bm\delta - \bm\gamma + \rho^{-1}\mathbf{u} \Vert_2^2 \right) ,
$$
with $f(\bm\delta)$ defined in \eqref{eq:f}. \\
The closed form update for $\hat{\bm\delta}$ is then obtained by taking the derivative of $f^*(\bm\delta, \bm\gamma,\mathbf{u})$ with respect to $\bm\delta$, and solving for \(\bm\delta\). Explicitly, we obtain:
$$
\frac{\partial f^*(\bm\delta, \bm\gamma,\mathbf{u})}{\partial \bm\delta} = \sum_{i = 1}^n \left[ \left(g_i^{\texttt{A}}\right)^2\bm\Omega\bm\delta - g_i^{\texttt{A}}\bm\Omega\mathbf{y}_i^c \right] + \rho\mathbf{D}^\top\left( \mathbf{D}\bm\delta - \bm\gamma\right) + \mathbf{D}^\top\mathbf{u} = 0 ,
$$
which, solving for $\bm\delta$ leads to the solution in \eqref{eq:deltaADMM}.

\subsection{Derivation for the $\hat{\bm\gamma}$ update} \label{appendixA2}
The update for $\hat{\bm\gamma}$ in \eqref{eq:gammaADMM} is obtained by minimizing \eqref{eq:augmLagrangian} with respect to $\bm\gamma$. As shown in \citet{zhu2017augmented}, this is equivalent to 
$$
\hat{\bm\gamma} =  \arg\min_{\bm\gamma} g^*(\bm\delta,\bm\gamma, \mathbf{u}) =  \arg\min_{\bm\gamma} \left( g(\bm\gamma) + \frac{\rho}{2} \Vert \mathbf{D}\bm\delta - \bm\gamma + \rho^{-1}\mathbf{u} \Vert_2^2 \right),
$$
with $g(\bm\gamma)$ defined as in \eqref{eq:g}. \\
The coordinate-wise closed form update for $\hat\gamma_j$ is therefore obtained as
\begin{eqnarray*}
\hat{\gamma}_j = \arg\min_{\gamma_j} g^{*}(\bm\delta, \gamma_j,\mathbf{u}) = \arg\min_{\gamma_j} \left( \lambda_\delta\sum_{j = 1}^p \vert \gamma_j \vert + \frac{\rho}{2} \left( \sum_{j = 1}^p \gamma_j^2 - 2\sum_{j = 1}^p \gamma_j d_j \right) - \sum_{j = 1}^p \gamma_j u_j \right) ,
\end{eqnarray*}
where $d_j$ is the $j$-th element of the vector $\mathbf{D}\bm\delta$. Since the problem is separable across coordinates, the closed form update for $\hat{\gamma}_j$ is obtained by taking the derivative of $g^{*}(\bm\delta, \gamma_j,\mathbf{u})$ with respect to $\gamma_j$, and solving for $\gamma_j$. Explicitly, we obtain:
$$
\frac{\partial  g^{*}(\bm\delta, \gamma_j,\mathbf{u})}{\partial \gamma_j} = \lambda_\delta s_{j,\gamma} + \rho \gamma_j - \rho d_j - u_j  = 0 ,
$$
where $\mathbf{s}_\gamma = (s_{1,\gamma}, \dots, s_{p,\gamma})$ denotes the subgradient of the $L_1$-norm with 
$$
s_{j,\gamma} = \begin{cases}
    \hspace{0.28cm} 1 \hspace{1.05cm} \text{if} \,\, \gamma_j > 0\\
    -1 \hspace{1.05cm} \text{if} \,\, \gamma_j < 0\\
    [-1, 1] \hspace{0.5cm} \text{if} \,\, \gamma_j = 0 .
\end{cases}
$$
Solving for $\gamma_j$, for $j = 1, \dots,p$, and considering the soft-thresholding operator as defined in Section \ref{sec:modelEst}, we obtain the solution in \eqref{eq:gammaADMM}.

\subsection{Derivation for the $\hat{\mathbf{u}}$ update} \label{appendixA3}
The closed-form update for the dual variable $\hat{\mathbf{u}}$ in \eqref{eq:dualvariableUpdate} follows directly from \citet{zhu2017augmented}.

\subsection{Derivation for the $\hat{\mathbf{g}}$ update} \label{appendixA4}
The update for $\hat{\mathbf{g}}^{\texttt{A}}$ in \eqref{eq:solutionG} is obtained by maximizing the penalized log-likelihood \eqref{eq:penLik} with respect to $g_i^{\texttt{A}}$, for $i = 1, \dots, n$. Specifically, we have
\begin{eqnarray*}
   \ell_p(\hat{\mathbf{g}}^{\texttt{A}} ,\bm\Theta) &\propto& - \frac{1}{2}\sum_{i = 1}^n (\mathbf{y}_i^c - \bm\delta g_i^\texttt{A})^\top \bm\Omega (\mathbf{y}_i^c - \bm\delta g_i^\texttt{A}) - \lambda_g\sum_{i = 1}^n \vert g_i^{\texttt{A}}\vert \\
   &\propto& -\frac{1}{2}\sum_{i = 1}^n \left((g_i^\texttt{A})^2 \bm\delta^\top\bm\Omega\bm\delta - 2g_i^\texttt{A}\bm\delta^\top\bm\Omega\mathbf{y}_i^c \right) - \lambda_g\sum_{i = 1}^n\vert g_i^\texttt{A}\vert .
\end{eqnarray*}
The closed form update for $g_i^{\texttt{A}}$ is obtained by taking the derivative of $\ell_p(\hat{\mathbf{g}}^{\texttt{A}} ,\bm\Theta)$ with respect to $g_i^{\texttt{A}}$, and solving for $g_i^{\texttt{A}}$. Explicitly, we obtain: 
\begin{eqnarray*}
   \frac{\partial \ell_p(\hat{\mathbf{g}}^{\texttt{A}} ,\bm\Theta)}{\partial g_i^{\texttt{A}}}  & = &  \bm\delta^\top \bm\Omega\mathbf{y}_i^c - g_i^{\texttt{A}}\bm\delta^\top\bm\Omega\bm\delta - \lambda_g s_{i,g} = 0 ,
\end{eqnarray*}
where $\mathbf{s}_g = (s_{1,g}, \dots, s_{n,g})$ denotes the subgradient of the $L_1$-norm with 
$$
s_{i,g} = \begin{cases}
    \hspace{0.28cm} 1 \hspace{1.05cm} \text{if} \,\, g_i > 0\\
    -1 \hspace{1.05cm} \text{if} \,\, g_i < 0\\
    [-1, 1] \hspace{0.5cm} \text{if} \,\, g_i = 0 .
\end{cases}
$$
Solving for $g_i$, for $i = 1, \dots,n$, and considering the soft-thresholding operator as defined in Section \ref{sec:modelEst}, we obtain the solution in \eqref{eq:solutionG}, which is subsequently modified with the additional thresholding step leading to \eqref{eq:thresholdG}. 

%--------------------------------------------------------------------------
\bibliographystyle{apalike}
\bibliography{biblio}

@article{schwarz1978estimating,
  title={Estimating the dimension of a model},
  author={Schwarz, Gideon},
  journal={The Annals of Statistics},
  volume = {6}, 
  number = {2},
  pages={461--464},
  year={1978},
  publisher={JSTOR}
}

@article{pritchard2000inference,
  title={Inference of population structure using multilocus genotype data},
  author={Pritchard, Jonathan K and Stephens, Matthew and Donnelly, Peter},
  journal={Genetics},
  volume={155},
  number={2},
  pages={945--959},
  year={2000},
  publisher={Oxford University Press}
}

@book{airoldi2014handbook,
  title={Handbook of mixed membership models and their applications},
  author={Airoldi, Edoardo M and Blei, David M and Erosheva, Elena A and Fienberg, Stephen E},
  year={2014},
  publisher={CRC press}
}

@book{bouveyron2019model,
  title={Model-based clustering and classification for data science: with applications in R},
  author={Bouveyron, Charles and Celeux, Gilles and Murphy, T Brendan and Raftery, Adrian E},
  year={2019},
  publisher={Cambridge University Press}
}

@book{mcnicholas2016mixture,
  title={Mixture model-based classification},
  author={McNicholas, Paul D},
  year={2016},
  publisher={CRC press}
}

@phdthesis{erosheva2002grade,
  title={Grade of membership and latent structure models with application to disability survey data},
  author={Erosheva, Elena Aleksandrovna},
  year={2002},
  school={Carnegie Mellon University}
}

@article{Zou2007a,
author = {Zou, Hui and Hastie, Trevor and Tibshirani, Robert},
issn = {0090-5364},
journal = {The Annals of Statistics},
number = {5},
pages = {2173--2192},
title = {{On the “degrees of freedom” of the lasso}},
volume = {35},
year = {2007}
}

@Manual{R,
    title = {R: A Language and Environment for Statistical Computing},
    author = {{R Core Team}},
    organization = {R Foundation for Statistical Computing},
    address = {Vienna, Austria},
    year = {2023},
    url = {https://www.R-project.org/},
  }

@Article{glmnetPack,
    title = {Regularization Paths for Generalized Linear Models via Coordinate Descent},
    author = {Jerome Friedman and Robert Tibshirani and Trevor Hastie},
    journal = {Journal of Statistical Software},
    year = {2010},
    volume = {33},
    number = {1},
    pages = {1--22},
    doi = {10.18637/jss.v033.i01},
  }

@article{casa2022group,
  title={Group-wise shrinkage estimation in penalized model-based clustering},
  author={Casa, Alessandro and Cappozzo, Andrea and Fop, Michael},
  journal={Journal of Classification},
  volume={39},
  number={3},
  pages={648--674},
  year={2022},
  publisher={Springer}
}

@article{lian2011shrinkage,
  title={Shrinkage tuning parameter selection in precision matrices estimation},
  author={Lian, Heng},
  journal={Journal of Statistical Planning and Inference},
  volume={141},
  number={8},
  pages={2839--2848},
  year={2011},
  publisher={Elsevier}
}

@article{fraley2002model,
  title={Model-based clustering, discriminant analysis, and density estimation},
  author={Fraley, Chris and Raftery, Adrian E},
  journal={Journal of the American statistical Association},
  volume={97},
  number={458},
  pages={611--631},
  year={2002},
  publisher={Taylor \& Francis}
}

@article{pan2007penalized,
  title={Penalized model-based clustering with application to variable selection.},
  author={Pan, Wei and Shen, Xiaotong},
  journal={Journal of Machine Learning Research},
  volume={8},
  number={5},
  year={2007}
}

@article{zhu2017augmented,
  title={An augmented ADMM algorithm with application to the generalized lasso problem},
  author={Zhu, Yunzhang},
  journal={Journal of Computational and Graphical Statistics},
  volume={26},
  number={1},
  pages={195--204},
  year={2017},
  publisher={Taylor \& Francis}
}

@inproceedings{heller2008statistical,
  title={Statistical models for partial membership},
  author={Heller, Katherine A and Williamson, Sinead and Ghahramani, Zoubin},
  booktitle={Proceedings of the 25th International Conference on Machine learning},
  pages={392--399},
  year={2008}
}

@article{blei2003latent,
  title={Latent dirichlet allocation},
  author={Blei, David M and Ng, Andrew Y and Jordan, Michael I},
  journal={Journal of Machine Learning Research},
  volume={3},
  pages={993--1022},
  year={2003}
}

@article{gormleyMurphy,
author = {Isobel Claire Gormley and Thomas Brendan Murphy},
title = {{A grade of membership model for rank data}},
volume = {4},
journal = {Bayesian Analysis},
number = {2},
pages = {265 -- 295},
year = {2009}
}

@article{dean2006using,
  title={Using unlabelled data to update classification rules with applications in food authenticity studies},
  author={Dean, Nema and Murphy, Thomas Brendan and Downey, Gerard},
  journal={Journal of the Royal Statistical Society Series C: Applied Statistics},
  volume={55},
  number={1},
  pages={1--14},
  year={2006},
  publisher={Oxford University Press}
}

@article{beer1852bestimmung,
  title={Bestimmung der Absorption des rothen Lichts in farbigen Fl{\"u}ssigkeiten},
  author={Beer, August},
  journal={Annalen der Physik},
  volume={162},
  number={5},
  pages={78--88},
  year={1852},
  publisher={Wiley Online Library}
}

@incollection{TEENTEH2014265,
title = {MEAT SPECIES DETERMINATION},
booktitle = {Encyclopedia of Meat Sciences},
publisher = {Academic Press},
edition = {2nd},
address = {Oxford},
pages = {265-269},
year = {2014},
author = {A.H. {Teen Teh} and G.A. Dykes},
}

@article{hassoun2020fraud,
  title={Fraud in animal origin food products: Advances in emerging spectroscopic detection methods over the past five years},
  author={Hassoun, Abdo and M{\aa}ge, Ingrid and Schmidt, Walter F and Temiz, Havva T{\"u}may and Li, Li and Kim, Hae-Yeong and Nilsen, Heidi and Biancolillo, Alessandra and A{\"\i}t-Kaddour, Abderrahmane and Sikorski, Marek and Sikorska, Ewa and Grassi, Silvia and Cozzolino, Daniel},
  journal={Foods},
  volume={9},
  number={8},
  pages={1069},
  year={2020},
  publisher={MDPI}
}

@article{casa2022parsimonious,
  title={Parsimonious Bayesian factor analysis for modelling latent structures in spectroscopy data},
  author={Casa, Alessandro and O’Callaghan, Tom F and Murphy, Thomas Brendan},
  journal={The Annals of Applied Statistics},
  volume={16},
  number={4},
  pages={2417--2436},
  year={2022},
  publisher={Institute of Mathematical Statistics}
}

@article{downey1997near,
  title={Near-and mid-infrared spectroscopies in food authentication: coffee varietal identification},
  author={Downey, Gerard and Briandet, Romain and Wilson, Reginald H and Kemsley, E Katherine},
  journal={Journal of Agricultural and Food Chemistry},
  volume={45},
  number={11},
  pages={4357--4361},
  year={1997},
  publisher={ACS Publications}
}

@article{connolly2006spectroscopic,
  title={Spectroscopic and Analytical Developments {L}td fingerprints brand spirits with ultraviolet spectrophotometry},
  author={Connolly, Christine},
  journal={Sensor Review},
  volume={26},
  number={2},
  pages={94--97},
  year={2006},
  publisher={Emerald Group Publishing Limited}
}

@article{kamal2015analytical,
  title={Analytical methods coupled with chemometric tools for determining the authenticity and detecting the adulteration of dairy products: A review},
  author={Kamal, Mohammad and Karoui, Romdhane},
  journal={Trends in Food Science \& Technology},
  volume={46},
  number={1},
  pages={27--48},
  year={2015},
  publisher={Elsevier}
}

@article{toher2007comparison,
  title={A comparison of model-based and regression classification techniques applied to near infrared spectroscopic data in food authentication studies},
  author={Toher, Deirdre and Downey, Gerard and Murphy, Thomas Brendan},
  journal={Chemometrics and Intelligent Laboratory Systems},
  volume={89},
  number={2},
  pages={102--115},
  year={2007},
  publisher={Elsevier}
}

@article{kelly2006application,
  title={Application of Fourier transform midinfrared spectroscopy to the discrimination between Irish artisanal honey and such honey adulterated with various sugar syrups},
  author={Kelly, J Daniel and Petisco, Cristina and Downey, Gerard},
  journal={Journal of Agricultural and Food Chemistry},
  volume={54},
  number={17},
  pages={6166--6171},
  year={2006},
  publisher={ACS Publications}
}

@article{downey1996authentication,
  title={Authentication of food and food ingredients by near infrared spectroscopy},
  author={Downey, Gerard},
  journal={Journal of Near Infrared Spectroscopy},
  volume={4},
  number={1},
  pages={47--61},
  year={1996},
  publisher={SAGE Publishing}
}

@article{reid2006recent,
  title={Recent technological advances for the determination of food authenticity},
  author={Reid, Linda M and O'Donnell, Colm P and Downey, Gerard},
  journal={Trends in Food Science \& Technology},
  volume={17},
  number={7},
  pages={344--353},
  year={2006},
  publisher={Elsevier}
}

@article{dimitrakopoulou2023does,
  title={Does traceability lead to food authentication? {A} systematic review from a European perspective},
  author={Dimitrakopoulou, Maria-Eleni and Vantarakis, Apostolos},
  journal={Food Reviews International},
  volume={39},
  number={1},
  pages={537--559},
  year={2023},
  publisher={Taylor \& Francis}
}

@article{murphy2010variable,
  title={Variable selection and updating in model-based discriminant analysis for high dimensional data with food authenticity applications},
  author={Murphy, Thomas Brendan and Dean, Nema and Raftery, Adrian E},
  journal={The Annals of Applied Statistics},
  volume={4},
  number={1},
  pages={396},
  year={2010},
  publisher={NIH Public Access}
}

@article{erosheva2007describing,
  title={Describing disability through individual-level mixture models for multivariate binary data},
  author={Erosheva, Elena A and Fienberg, Stephen E and Joutard, Cyrille},
  journal={The Annals of Applied Statistics},
  volume={1},
  number={2},
  pages={346},
  year={2007}
}

@article{gruhl2013tale,
  title={A tale of two (types of) memberships: Comparing mixed and partial membership with a continuous data example},
  author={Gruhl, Jonathan and Erosheva, Elena A and Ghahramani, Z and Mohamed, S and Heller, K A},
  journal={Handbook of Mixed Membership Models and Their Applications},
  publisher={CRC press},
  pages={15--38},
  year={2014}
}

@Book{whittaker:1990,
  author = {Whittaker, J.},
  title = {Graphical Models in Applied Multivariate Statistics},
  publisher = {Wiley},
  year = {1990},
}

@article{friedman:etal:2008,
  title        = {Sparse inverse covariance estimation with the graphical lasso},
  author       = {Friedman, Jerome and Hastie, Trevor and Tibshirani, Robert},
  year         = 2008,
  journal      = {Biostatistics},
  volume       = 9,
  number       = 3,
  pages        = {432--441}
}

@article{gen_lasso,
author = {Ryan J. Tibshirani and Jonathan Taylor},
title = {{The solution path of the generalized lasso}},
volume = {39},
journal = {The Annals of Statistics},
number = {3},
publisher = {Institute of Mathematical Statistics},
pages = {1335 -- 1371},
year = {2011}
}

@article{tibshirani2005sparsity,
  title={Sparsity and smoothness via the fused lasso},
  author={Tibshirani, Robert and Saunders, Michael and Rosset, Saharon and Zhu, Ji and Knight, Keith},
  journal={Journal of the Royal Statistical Society Series B: Statistical Methodology},
  volume={67},
  number={1},
  pages={91--108},
  year={2005},
  publisher={Oxford University Press}
}

@article{tibshirani1996regression,
  title={Regression shrinkage and selection via the lasso},
  author={Tibshirani, Robert},
  journal={Journal of the Royal Statistical Society Series B: Statistical Methodology},
  volume={58},
  number={1},
  pages={267--288},
  year={1996},
  publisher={Oxford University Press}
}

@article{kim_TF,
author = {Kim, Seung-Jean and Koh, Kwangmoo and Boyd, Stephen and Gorinevsky, Dimitry},
title = {$\ell_1$ Trend Filtering},
journal = {SIAM Review},
volume = {51},
number = {2},
pages = {339-360},
year = {2009}
}

@article{trend_filtering,
author = {Ryan J. Tibshirani},
title = {{Adaptive piecewise polynomial estimation via trend filtering}},
volume = {42},
journal = {The Annals of Statistics},
number = {1},
pages = {285 -- 323},
year = {2014}
}

@article{ferraccioli2023adaptive,
  title={An adaptive functional regression framework for locally heterogeneous signals in spectroscopy},
  author={Ferraccioli, Federico and Casa, Alessandro and Stefanucci, Marco},
  journal={Journal of the Royal Statistical Society Series C: Applied Statistics},
  volume={73},
  number={5},
  pages={1370--1388},
  year={2024},
  publisher={Oxford University Press UK}
}

@article{boydADMM,
    year = {2011},
    volume = {3},
    journal = {Foundations and Trends® in Machine Learning},
    title = {Distributed Optimization and Statistical Learning via the Alternating Direction Method of Multipliers},
    number = {1},
    pages = {1-122},
    author = {Stephen Boyd and Neal Parikh and Eric Chu and Borja Peleato and Jonathan Eckstein}
}

@article{ramdasADMM,
    author = {Aaditya Ramdas and Ryan J. Tibshirani},
    title = {Fast and Flexible {ADMM} Algorithms for Trend Filtering},
    journal = {Journal of Computational and Graphical Statistics},
    volume = {25},
    number = {3},
    pages = {839-858},
    year  = {2016}
}

@article{kelly2006potential,
  title={Potential of near infrared transflectance spectroscopy to detect adulteration of Irish honey by beet invert syrup and high fructose corn syrup},
  author={Kelly, J Daniel and Petisco, Cristina and Downey, Gerard},
  journal={Journal of Near infrared spectroscopy},
  volume={14},
  number={2},
  pages={139--146},
  year={2006},
  publisher={SAGE Publishing}
}

@article{hineno1977infrared,
  title={Infrared spectra and normal vibration of $\beta$-d-glucopyranose},
  author={Hineno, Masakazu},
  journal={Carbohydrate Research},
  volume={56},
  number={2},
  pages={219--227},
  year={1977},
  publisher={Elsevier}
}

@article{vasko1972infrared,
  title={Infrared and raman spectroscopy of carbohydrates.: Part {II}: Normal coordinate analysis of $\alpha$-D-glucose.},
  author={Vasko, PD and Blackwell, J and Koenig, JL},
  journal={Carbohydrate Research},
  volume={23},
  number={3},
  pages={407--416},
  year={1972},
  publisher={Elsevier}
}

@article{saeys2008potential,
  title={Potential applications of functional data analysis in chemometrics},
  author={Saeys, Wouter and De Ketelaere, Bart and Darius, Paul},
  journal={Journal of Chemometrics},
  volume={22},
  number={5},
  pages={335--344},
  year={2008},
  publisher={Wiley Online Library}
}

@article{marco2024functional,
  title={Functional mixed membership models},
  author={Marco, Nicholas and {\c{S}}ent{\"u}rk, Damla and Jeste, Shafali and DiStefano, Charlotte and Dickinson, Abigail and Telesca, Donatello},
  journal={Journal of Computational and Graphical Statistics},
  volume={33},
  number={4},
  pages={1139--1149},
  year={2024},
  publisher={Taylor \& Francis}
}

@inproceedings{casa2023partial,
  title={Partial membership models for high-dimensional spectroscopy data},
  author={Casa, Alessandro and Murphy, Thomas Brendan and Fop, Michael},
  booktitle={Cladag 2023 Book of Abstracts and Short Papers. 14th Scientific Meeting of the Classification and Data Analysis Group},
  pages={82--85},
  year={2023},
  organization={Pearson}
}

@article{codazzi2022gaussian,
  title={Gaussian graphical modeling for spectrometric data analysis},
  author={Codazzi, Laura and Colombi, Alessandro and Gianella, Matteo and Argiento, Raffaele and Paci, Lucia and Pini, Alessia},
  journal={Computational Statistics \& Data Analysis},
  volume={174},
  pages={107416},
  year={2022},
  publisher={Elsevier}
}

@article{hou-liu2022, 
AUTHOR = {Hou-Liu, Jason and Browne, Ryan P.},
TITLE={Chimeral clustering},
JOURNAL={Journal of Classification},
VOLUME={39}, 
PAGES={171--190},
YEAR={2022}
}

@book{sun2009infrared,
  title={Infrared Spectroscopy for Food Quality Analysis and Control},
  author={Sun, Da-Wen},
  year={2009},
  publisher={Academic press}
}

@article{Fop2022,
author={Fop, Michael
and Mattei, Pierre-Alexandre
and Bouveyron, Charles
and Murphy, Thomas Brendan},
title={Unobserved classes and extra variables in high-dimensional discriminant analysis},
journal={Advances in Data Analysis and Classification},
year={2022},
month={Mar},
day={01},
volume={16},
number={1},
pages={55-92},
issn={1862-5355},
doi={10.1007/s11634-021-00474-3},
url={https://doi.org/10.1007/s11634-021-00474-3}
}

\end{document}